\documentclass[twocolumn]{aastex631}

\shorttitle{}
\shortauthors{FU, HUANG, $\&$ YU}

\usepackage{amsmath} 
\usepackage{graphicx}
\usepackage{booktabs} 
\usepackage{tabularx}
\usepackage{mathrsfs} 
\usepackage{subfigure}
\usepackage{CJK}
\usepackage[mathscr]{eucal}
\graphicspath{{./}{figures/}}
\begin{document}
\begin{CJK*}{UTF8}{gbsn}

\title{Boundary Layers of Circumplanetary Disks around Spinning Planets \\ II. Global Modes with Azimuthal Magnetic Fields}

\correspondingauthor{Cong Yu}
 \email{yucong@mail.sysu.edu.cn}
\author[0000-0002-9108-9500]{Zhihao Fu (付智豪)}
\affiliation{School of Physics and Astronomy, Sun Yat-Sen University, Zhuhai 519082, China}
\affiliation{CSST Science Center for the Guangdong-Hong Kong-Macau Greater Bay Area, Zhuhai 519082, China}
\affiliation{State Key Laboratory of Lunar and Planetary Sciences, Macau University of Science and Technology, Macau, China}
\affiliation{Department of Physics, University of Hong Kong, Hong Kong SAR, China}
\author[0000-0002-7276-3694]{Shunquan Huang (黄顺权)}
\affiliation{School of Physics and Astronomy, Sun Yat-Sen University, Zhuhai 519082, China}
\affiliation{CSST Science Center for the Guangdong-Hong Kong-Macau Greater Bay Area, Zhuhai 519082, China}
\affiliation{State Key Laboratory of Lunar and Planetary Sciences, Macau University of Science and Technology, Macau, China}
\affiliation{Department of Physics and Astronomy, University of Nevada, Las Vegas, 4505 South Maryland Parkway, Las Vegas, NV 89154, USA}
\author[0000-0003-0454-7890]{Cong Yu (余聪)}
\affiliation{School of Physics and Astronomy, Sun Yat-Sen University, Zhuhai 519082, China}
\affiliation{CSST Science Center for the Guangdong-Hong Kong-Macau Greater Bay Area, Zhuhai 519082, China}
\affiliation{State Key Laboratory of Lunar and Planetary Sciences, Macau University of Science and Technology, Macau, China}




\begin{abstract}
The accretion of material from disks onto weakly magnetized objects invariably involves its traversal through a material surface, known as the boundary layer (BL). 
Our prior studies have revealed two distinct global wave modes for circumplanetary disks (CPDs) with BLs exhibit opposite behaviors in spin modulation. 
We perform a detailed analysis about the effect of magnetic fields on these global modes, highlighting how magnetic resonances and turning points could complicate the wave dynamics. The angular momentum flux becomes positive near the BL with increasing magnetic field strength. 
We also examine the perturbation profile to demonstrate the amplification of magnetic fields within the BL. 
The dependence of growth rates on the magnetic field strength, and the spin rate are systematically investigated. 
We find that stronger magnetic fields tend to result in lower terminal spin rates. 
We stress the potential possibility for the formation of angular momentum belts and pressure bumps. 
The implication for the spin evolution and quasi-period oscillations observed in compact objects are also briefly discussed. 
Our calculations advance the understanding of magnetohydrodynamical (MHD) accretion processes  and lays a foundation for observational
studies and numerical simulations.
\end{abstract}

% \keywords{accretion, accretion disks  - planet formation - hydrodynamics - waves - instabilities}


% \input{Sec/sec1}
% \input{Sec/sec2}
% \input{Sec/sec3}
% \input{Sec/sec3-1}
% \input{Sec/sec4}
% \input{Sec/sec5}
% \input{Sec/sec6}
% \input{Sec/sec7}
% \input{Sec/sec8}
% \input{Sec/Appendix}
\section{Introduction}
Accretion usually occurs in a system consisting of a disk surrounding a central object, such as protoplanets, protostars, white dwarfs/neutron stars in binaries, active galactic nuclei, or black holes. 
\textcolor{black}{In the presence of strong magnetic fields, accreting materials follow trajectories along magnetic field lines toward the central object.}
% During the magnetospheric accretion in the context of strong magnetic fields, materials follow trajectories along magnetic field lines toward the central object.
However, in situations characterized by weakly magnetized or purely hydrodynamic backgrounds, accretion experiences a transition from rapid Keplerian rotation to a slower, rigid rotation, which is known as the $\textit{boundary layer}$ (BL)\footnote{With the exclusion of the black hole, as it lacks a material surface for connection with the disk.}. 
Young stellar objects of the FU Ori type \citep{1993ApJ...415L.127P} and cataclysmic variables \citep{1978A&A....63..265K,1993Natur.362..820N} are expected to form BLs. Additionally, a similar mechanism, involving the presence of a spreading layer, also operates in neutron stars of low-mass X-ray binaries \citep{2016ApJ...817...62P}.
\begin{figure*}[t]
    \centering
    \includegraphics[width=0.75\linewidth]{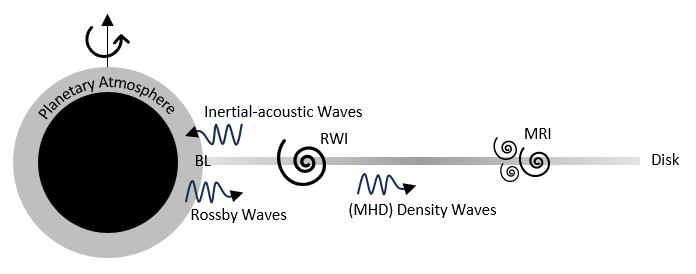}
    \caption{Illustration of waves and instabilities in a planet-disk system. 
    The BL is located at the interface between the planet and the surrounding disk, marking the transition from Keplerian rotation to rigid-body rotation. 
    Waves are generated within this region as inertial-acoustic waves propagating into the planet and MHD density waves propagating into the disk.
    Simultaneously, a subset of Rossby waves are generated inside the planetary atmosphere and propagate outward due to planetary rotation.
    At the inner edge of the disk, significant variations in the angular velocity within the BL promote the development of the RWI in the vicinity of the planet. 
    In addition, the evolution of the magnetic field can give rise to the MRI inside the disk.
    }
    \label{WaI}
\end{figure*}
Recent numerical simulations have uncovered a number of phenomena that deviate from the conventional model employing turbulent viscosity. \cite{2013ApJ...770...67B} identified non-axisymmetric global acoustic modes in hydrodynamic simulations, suggesting that acoustic waves generated within the BL could transfer angular momentum to both the planet and the disk. Subsequent magnetohydrodynamic (MHD) simulations confirm the magnetic field amplification within the BL \citep{2013ApJ...770...68B}, and the formation of angular momentum belt around planets could significantly reduce the transport efficiency \citep{2017ApJ...835..238B}. Furthermore, \cite{2022MNRAS.509..440C,2022MNRAS.512.2945C} investigated the characteristics of these wave modes and discovered novel vortex-driven modes near the BL.

Prior studies predominantly concentrate on problems of non-rotating planets. 
In fact, central objects typically possess a certain degree of rotation, which introduces complexity to the dynamical processes of the system. 
In this paper, our main focus is on addressing the problem of thin disks around rotating planets characterized by the BL. 
Figure \ref{WaI} roughly depicts the waves and instabilities in a  magnetized planet-disk system. 
Due to the sharp supersonic variation in angular velocities, sonic instabilities can result in the excitation of acoustic waves within the BL \citep{2012ApJ...752..115B}. 
These waves subsequently propagate in both directions and manifest themselves within the planetary atmosphere as inertial-acoustic waves. 
On the other hand, when viewed from a reference frame that co-rotates with the planet, a new branch of low-frequency waves emerges, referred to as Rossby waves. These waves originate from the interior and mostly propagate outwards \citep{2023ApJ...945..165F}.

To account for the loss of angular momentum during the accretion process from disks to planets, magnetorotational instability (MRI) stands out as a pivotal transport mechanism. The development of this instability necessitates a decreasing angular velocity with respect to the radius, $d\Omega/dr<0$. 
However, this condition is violated within the boundary layer (BL) where $d\Omega/dr>0$, thereby preventing the initiation of MRI.
\textcolor{black}{Additionally, by calculating the spatial distribution of the ionization degree in the disk, \citet{2014ApJ...785..101F} found that it is hard for the MRI turbulence to develop well in CPDs except for the thin active layer at the disk surface. 
Due to the lack of angular momentum transfer through the MRI, gas is more likely to pile up until gravitational instability occurs.
However, we can see that such an initial distribution of angular velocity is unstable to the Rossby wave instability (RWI), leading to the formation of vortices and the spiral arms in disks \citep{1999ApJ...513..805L}.
It is possible for accretion to proceed without the MRI turbulence in CPDs.
}
% Nevertheless, it is noteworthy that such an initial distribution of angular velocity is unstable to the Rossby wave instability (RWI), leading to the formation of vortices and the spiral arms in disks \citep{1999ApJ...513..805L}.

A deeper understanding of the behavior of global modes related to BLs would be of great significance. 
Theoretical analysis can provide us the insights into the underlying physical mechanisms behind the phenomena.
Through local linear analysis, one can assess the stability by examining the imaginary part of eigenvalues. 
With the global analysis, we can find that these waves and instabilities form a coherent pattern displaying the same angular velocity $\Omega_{\rm P}$.
In \cite{2023ApJ...945..165F} (hereafter Paper I), we identified two types of modes in hydrodynamics, named inertial-acoustic modes ($p$-modes) and Rossby modes ($r$-modes). 
The major difference between these two modes is the angular momentum flux (accretion rate) inside the planet. The $p$-modes exhibit negative (positive) values, while the $r$-modes exhibit positive (negative) values. 
Thus, $p$-modes can lead to gas accretion and speed up planets but $r$-modes can shed up mass and slow down planets
\footnote{Note that the exact expression for the accretion rate also contains a second-order perturbation of velocities. Our analysis has only been carried out up to the first order, so there are some errors.}.
Within the BL, there is a negative angular momentum and a positive accretion rate, which implies that a gap in density would form around the planet. 
In the disk, there is a positive angular momentum and a negative accretion rate, so waves speed up materials and result in the expansion of disks.
Besides, we show that the transition in the dominance of $p$ and $r$ modes marks the terminal spin, so a planet could not spin up to near break-up velocities. 
Our work is in good agreement with numerical simulations \citep{1993PhDT........41P, 2021ApJ...921...54D} and gives a possible explanation for the low spin rate observed in planetary companions and brown dwarfs \citep{2018NatAs...2..138B}.

When the magnetic field is introduced, new resonances appear, such as magnetic resonance and Alfv{\'e}n resonance \citep{2015ApJ...802...54U,2003MNRAS.341.1157T,2008ApJ...679..813M}. These resonances and their turning points can cause a change to the original configuration of the propagable and evanescent regions of waves.
Additionally, the density wave in the disk splits into fast and slow MHD density waves, making it possible for new magnetosonic modes to emerge \citep{1998ApJ...493..102L}.  
In this paper, we focus on the effect of azimuthal magnetic fields on global modes.

Our paper is organized as follows. 
In $\S$\ref{Sec2} we obtain the perturbation equations and introduce corresponding solving methods.
In $\S$\ref{Sec3} we analyze the waves and instabilities present in the system.
In $\S$\ref{BLMoSP} we delve into the properties of boundary layer modes.
In $\S$\ref{Impact_Mag} we study the impact of magnetic field strength on global modes.
In $\S$\ref{Sec5} we compare our theory with numerical simulations and discuss the implications.
In $\S$\ref{Sec6} we summarize our conclusions.

\section{Basic theory}\label{Sec2}
\subsection{Fundamental equations}
The ideal magnetohydrodynamic (MHD) equations for compressible fluids can be written as follows
\begin{equation}\label{mass}
\frac{\partial \rho}{\partial t}+\nabla \cdot (\rho \boldsymbol{v})=0 \ , \
\end{equation}
\begin{equation}\label{motion}
\frac{\partial \boldsymbol{v}}{\partial t}+
\boldsymbol{v}\cdot \nabla \boldsymbol{v}
=-\frac{1}{\rho}\nabla P-
\nabla \Phi+
\frac{1}{\rho}(\nabla \times \boldsymbol{B})\times \boldsymbol{B} \ , \
\end{equation}
\begin{equation}
\frac{\partial \boldsymbol{B}}{\partial t}=
\nabla \times (\boldsymbol{v}\times \boldsymbol{B}) \ , \
\end{equation}
\begin{equation}\label{energy}
\frac{\partial P}{\partial t}+
\boldsymbol{v}\cdot\nabla P+
\gamma P \nabla \cdot \boldsymbol{v}=0  \ , \ 
\end{equation}
where $\rho, \boldsymbol{v}, P, \Phi, \boldsymbol{B}$ and $\gamma$ are the density, velocity, pressure, gravitational potential, magnetic field, and adiabatic index, respectively.
Note that the Lorentz force term is written in a more compact form by absorbing the coefficient $4\pi$ into the magnetic field.
\subsection{Initial Physical Setup}\label{ISet}
For the analysis of the planet-disk system, we adopt cylindrical coordinates $(r, \phi, z)$. 
Due to the substantial disparity in mass between the disk and the planet, the influence of CPD self-gravity can be disregarded. In this study, we exclude vertical stratification and instead employ a cylindrically-symmetric potential 
\begin{equation}
    \Phi=-\frac{1}{r} \label{potential} \ .
\end{equation}

In our calculations, the initial velocity profile is characterized by circular motion, $\boldsymbol{v} = (0, r\Omega(r), 0)$, where $\Omega(r)$ can be divided into three distinct regions.  
Within the disk, the gas is expected to exhibit Keplerian rotation with a power-law dependence of $r^{-3/2}$, which is governed by the central gravitational potential.
For the inner region of the planet below the atmosphere, we treat it as a rigid body with a constant rotation rate $\Omega_0$.
During the process of accretion from the disk to the planet, a significant decrease in the rotation profile occurs, referred to as the boundary layer (BL). 
% To account for this, we use a linear function to represent the variation in angular velocity within the BL while ensuring a smooth transition in our calculations.
Here we use a linear function to represent the variation in angular velocity within the BL. 
Mathematically, these functions can be written as
\begin{small}
\begin{equation}\label{Omega_Profile}
\Omega(r)=\left \{
\begin{array}{lcl}
\Omega_0 & & 
{r < r_* - \delta r}\\
r^{-3/2}& & 
{r >r_* + \delta r}\\
\Omega_0 + \left[(r_*+ \delta r)^{-3/2}-\Omega_0 \right]
\frac{r+\delta r - r_*}{2 \delta r}& & 
{r_* - \delta r \leq r}\\
&&  {\leq r_* + \delta r} \ ,
\end{array} \right.
\end{equation}
\end{small}
where $\delta r$ is the width of the BL. and $r_* = 1$ is the surface of the planet. 
\textcolor{black}{To ensure a smooth transition within the BL, we apply the average method where each data point is averaged with a number of adjacent points on either side.
This approach makes the smoothing length dependent on the resolution of the calculation. We discuss the effect of resolution in Appendix \ref{Resolution}, where the results indicate that the solution converges at high resolution. Hence, we use the resolution at which the solutions converge.}
% When it comes to the initial distribution of the magnetic field, we assume that the magnetic field is confined to the disk. It is primarily characterized by an azimuthal component denoted as $\boldsymbol{B} = (0, B_{\phi}(r), 0)$, where

\textcolor{black}{Regarding the initial distribution of the magnetic field, it is primarily characterized by an azimuthal component denoted as $\boldsymbol{B} = (0, B_{\phi}(r), 0)$, where}
\begin{equation}\label{imag}
    B_{\phi} = \lambda B_0 r^{-\alpha}
\end{equation}
% for $r > r_* + \delta r$.
Here, $B_0 = \sqrt{p_*}$ and $p_*$ corresponds to the pressure at $r_*$. 
The power index $\alpha$ is commonly chosen as 1.
The parameter $\lambda$ controls the strength of the magnetic field.
% \textcolor{black}{Since most of previous numerical simulations mainly studied the case of unmagnetized planets, in this paper, we will also focus on this case where the magnetic field is confined to the disk (for $r > r_* + \delta r$). 
% Furthermore, we will consider the case where the magnetic field extends to the interior of the planet, which to some extent corresponds to the case of magnetized planets (see Appendix \ref{AppendixB}).
% }

To distinguish between different astrophysical systems, we will introduce two dimensionless parameters defined at $r_* = 1$. 
Since $V(r_*) = 1$, the characteristic Mach number is given by 
\begin{equation}
    \mathcal{M} \equiv V(r_*)/c_s = c_s^{-1},
\end{equation}
where $c_s$ is the sound speed.
The Mach number exhibits a significant correlation with the scale height of disks, denoted as $h$, as well as that of planetary atmospheres, denoted as $H$. 
The vertical equilibrium of Keplerian disks implies that $h \sim c_s/\Omega$.
Moreover, the equilibrium of an isothermal stratified atmosphere suggests that $H = c_s^2 /g$. At $r_* = 1$, the corresponding values are $h = M^{-1}$ and $H = M^{-2}$, resulting in the relation $H = h / M \ll h$.
The magnitude of the Mach number can be determined through observational measurements or numerical simulations. For instance, circumplanetary disks typically exhibit $\mathcal{M} \sim 2-3$ \citep{2009MNRAS.397..657A}.

We define the plasma parameter as
\begin{equation}
    \beta \equiv \frac{ P }{1/2 B_{\phi}^2} =
    \frac{c^2_s}{c^2_a},
\end{equation}
which quantifies the relative significance of magnetic pressure in comparison to gas pressure. 
While $\beta$ generally varies with radius, we specifically select its value at $r_*$ as our designated parameter, where $\beta = 2/\lambda^2$.

\subsection{Equilibrium State}
From equation \eqref{motion} we can obtain the equilibrium state 
\begin{equation}\label{equlibrium}
    \frac{V^2}{r}
    \equiv 
    \Omega^2 r
    = 
    \frac{1}{\rho} 
    \left[
    \frac{d }{d r} \left(P + \frac{1}{2}B_{\phi}^2\right)
    + \frac{B_{\phi}^2}{r} 
    \right]
    +\frac{d\Phi}{dr}.
\end{equation}
Then we can define the effective gravitational acceleration revised by the centrifugal force, namely $g \equiv \Omega^2 r -d \Phi/dr$. 

In order to solve the initial density profile, we substitute equation \eqref{potential} and \eqref{imag} into $\eqref{equlibrium}$ and take into account the relation $P = \gamma^{-1} \rho c_s^2$, where $c_s$ is a constant. This results in the following expression 
\begin{equation}
    \gamma^{-1} c_s^2 \frac{d \rho }{d r} = \rho \left(
    \Omega^2 r - \frac{1}{r^2}
    \right)
    - \left( 1 - \alpha \right)
    \frac{B_{\phi}^2 }{r}.
\end{equation}
It is worth noting that the second magnetic term on the right-hand would vanish if we assume $\alpha =1$. Hence, the introduction of magnetic fields has no influence on the density profile.
Likewise, the first term also vanishes in a disk where $\Omega = r^{-3/2}$, which means a constant density across the disk.
However, the first term becomes negative in the planetary atmosphere, indicating a decrease in density as the radius increases. 
Higher values of $\mathcal{M}$ lead to steeper density profiles, while larger $\Omega_0$ yields less steep profiles.
By integrating the equation, it can be derived that the density $\rho$ is exponentially proportional to $\exp( \mathcal{M}^2 /r)$ within the planet for $\Omega_0 = 0$. 

\begin{figure}[t]
    \centering
    \includegraphics[width=1\linewidth]{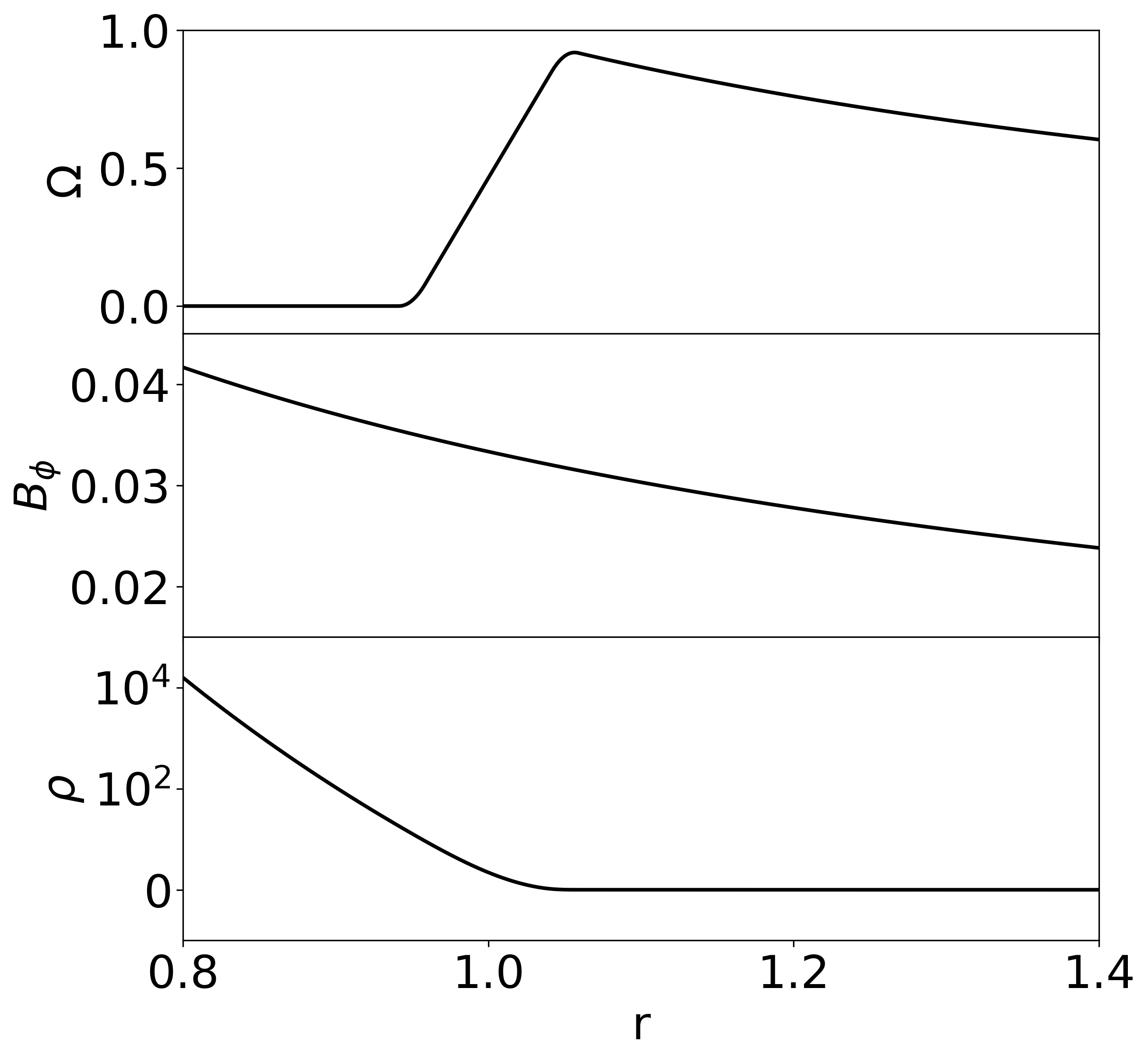}
    \caption{Initial radial profiles for angular velocity $\Omega$, magnetic field $B_{\phi}$ and density $\rho$.}
    \label{Equ}
\end{figure}

\textcolor{black}{Figure \ref{Equ} illustrates the initial profile of angular velocity $\Omega(r)$ defined by the equation \eqref{Omega_Profile} with parameters $\Omega_0=0$ and $\delta r =0.05$, and the initial profile of magnetic fields $B_{\phi}(r)$ defined by the equation \eqref{imag} with $\lambda=0.2$ and $\mathcal{M}=6$. The exact density profile is resolved by assuming that $\gamma = 1$ and $\rho = 1$ in the disk. }

\subsection{Perturbation Analysis}
We use the prefix $\delta$ to denote the perturbation of any physical quantity $f$. For non-axisymmetric structures, it can be expanded in a Fourier series as $\delta f\sim f(r)\exp(im\theta+ik_{z}z-i\omega t)$. By defining $\Psi = (\delta P + B_{\phi}\delta B_{\phi})/\rho$, the basic MHD equations can be linearized as follows
\begin{equation}\label{dmass}
i\sigma \frac{\delta\rho}{\rho}
=\left(\frac{1}{L_{\rho}}+
\frac{1}{r}\right)\delta v_{r}+
\frac{\partial \delta v_{r}}{\partial r}+
ik_{\phi} \delta v_{\phi}+
ik_{z} \delta v_{z} \ , \
\end{equation}
\[
-i\sigma v_r - 2 \Omega v_{\phi } =
- \frac{\partial \Psi}{\partial r}
- \frac{\Psi}{L_{\rho}} + \frac{\delta \rho}{\rho} g
\]
\begin{equation}
\qquad \qquad \qquad + \frac{ik_{\phi }B_{\phi}}{\rho} \delta B_r 
- \frac{2 B_{\phi}}{\rho r} \delta B_{\phi} \ , \
\end{equation}
%\begin{equation}
%-i\sigma v_{\phi} + \frac{\kappa^2}{2\Omega}v_r = 
%- i k_{\phi} \Psi 
%+ \frac{1}{\rho} \left( \frac{d B_{\phi} }{d r } +\frac{B_{\phi}}{r}\right) \delta B_{r}
%+ \frac{1}{\rho}ik_{\phi}B_{\phi} \delta B_{\phi}
%\end{equation}
\begin{equation}
\frac{\kappa^2}{2\Omega}v_r -i\sigma v_{\phi} = 
\frac{d (rB_{\phi}) }{\rho r d r } \delta B_{r}
- i k_{\phi} \Psi + \frac{ik_{\phi}B_{\phi}}{\rho} \delta B_{\phi} \ , \ 
\end{equation}
\begin{equation}
-\sigma v_z  = 
- k_z \Psi
+ \frac{1}{\rho} k_{\phi}B_{\phi} \delta B_z \ , \ 
\end{equation}
\begin{equation}
-\sigma \delta B_{r}=
k_{\phi}B_{\phi}\delta v_{r} \ , \
\end{equation}
\[
-i\sigma \delta B_{\phi}=
-\frac{d B_{\phi}}{d r}\delta v_{r} - ik_{z}B_{\phi}\delta v_{z} 
\]
\begin{equation}
\qquad \qquad \qquad \qquad + \frac{d\Omega}{d\ln r}\delta B_{r} - B_{\phi}\frac{\partial \delta v_{r}}{\partial r} \ , \
\end{equation}
\begin{equation}
-\sigma \delta B_{z}=
k_{\phi}B_{\phi}\delta v_{z} \ , \ 
\end{equation}
\begin{equation}\label{denergy}
 \Psi=
 \frac{1}{\rho}B_{\phi}\delta B_{\phi}
+  c_{s}^{2} \frac{\delta \rho}{\rho}
+\frac{ c_{s}^{2}}{i\sigma  L_{s}}\delta v_{r} \ . \
\end{equation}
The local Doppler-shifted frequency is given by $\sigma = \omega - m \Omega = m(\Omega_{\rm P} - \Omega)$, where $m$ is the mode number. The azimuthal wave number is defined as $k_{\phi}=m / r$, and the sound velocity is given by $c_{s}^{2}=\gamma P / \rho$.
\begin{equation}\label{epicyclic}
    \kappa^2\equiv \frac{1}{r^3}\frac{d(\Omega^2 r^4)}{d r}=
    4 \Omega^2 + 2 \Omega \frac{d \Omega}{d r}r \ ,  
\end{equation}
is the square of radial epicyclic frequency. In addition, we introduce the length scale of entropy variation $L_{\rm S}$ which satisfies
\begin{equation}
    \frac{1}{L_{\rm S}}= \frac{1}{\gamma}\frac{d}{dr} \left[\ln \left( \frac{P}{\rho^{\gamma}} \right) \right] \ .
\end{equation}
Note that in the adiabatic process, the spatial scale $L_{\rm S}$ tends to infinity. Similarly, the length scale of pressure, density, and magnetic field can be written as
\begin{equation}
    \frac{1}{L_{\rm P}}=\frac{1}{\gamma P}  \frac{d P}{d r}  \ ,  \
    \frac{1}{L_{\rho}}=\frac{1}{\rho}  \frac{d \rho }{d r} \ ,  \
    \frac{1}{L_m} = \frac{1}{B_{\phi}}\frac{d B_{\phi}}{d r},
\end{equation}
respectively. Note that $1/L_{\rm S}$=$1/L_{\rm P}$ $- 1/L_{\rho}$. The above equation \eqref{dmass}-\eqref{denergy} can be reduced to two first-order linear equations
\begin{equation}
\label{leq1}
\frac{\partial \delta v_{r}}{\partial r}+
A_{11}\delta v_{r}+A_{12}\Psi=0
\end{equation}
\begin{equation}
\label{leq2}
\frac{\partial \Psi}{\partial r}+
A_{21}\delta v_{r}+A_{22}\Psi=0
\end{equation}
where
\begin{align}
A_{11} &= \frac{1}{(\sigma^2 - \sigma^2_M)} \left[ \frac{\sigma^2}{L_1} + \frac{ c_s^2}{(c_a^2+c_s^2)}\frac{k_{\phi}\kappa^2 }{2\Omega}\sigma 
+ \frac{\sigma_M^2}{r} 
\right. \nonumber \\
& \quad  \left. 
+\frac{ 1}{(c_a^2+c_s^2)} \frac{d\Omega}{d \ln r} \frac{k_{\phi}c_a^2}{\sigma}\left( \sigma^2 - k_{\phi}^2c_s^2 \right) \right]
,
\end{align}
\begin{equation}
    A_{12} = - i\sigma \left[
\frac{1}{c_a^2+c_s^2}\frac{\sigma^2 - k_{\phi}^2c_s^2}{\sigma^2 - \sigma^2_M}
- \frac{k_z^2}{\sigma^2 - k_{\phi}^2c_a^2}
\right],
\end{equation}
\begin{align}
    A_{21} &= 
- i \frac{1}{\sigma}
\left[
\sigma^2
- \kappa^2
- k_{\phi}^2c_a^2 
+ \frac{g}{L_S} 
- \frac{2\Omega k_{\phi}}{\sigma}(\frac{c_a^2}{r} 
+ \frac{c_a^2}{L_m})
\right. \nonumber \\
& \quad \left.
+ \frac{g_m}{L_m}
+ g_m\frac{k_{\phi}}{\sigma} \frac{d\Omega}{d \ln r}
-g_m A_{11}
\right]
\end{align}
\begin{equation}
\label{coe22}
    A_{22} = -\frac{2\Omega  k_{\phi}}{\sigma } - \frac{1}{L_S}
+ \frac{g_m}{c_a^2+c_s^2}\frac{\sigma^2 - k_{\phi}^2c_s^2}{\sigma^2 - \sigma^2_M},
\end{equation}
where
$$
    c_a ^2 = \frac{B^2_{\phi}}{\rho} \ , \ 
    \sigma_M^2 = \frac{k_{\phi}^2 c_s^2c_a^2 }{c_s^2 + c_a^2} \ , 
$$
$$
\frac{1}{L_1} =
        \frac{ c_s^2}{(c_a^2+c_s^2)}
         \left[
        \frac{1}{r} + \frac{1}{L_{\rho}} + \frac{1}{L_S} + \frac{c^2_a}{c_s^2}\frac{1}{L_m}
        \right],
$$
and 
$$
    g_m = \frac{2\Omega  k_{\phi}c_a^2}{\sigma } + \frac{2 c_a^2}{r} + \frac{c_a^2}{c_s^2}g .
$$
Disregarding the magnetic terms, the equations degenerate into the hydrodynamics form discussed in Paper I. 
Moreover, by combining equation \eqref{leq1} and \eqref{leq2}, we derive a second-order differential equation for $\Psi$, which reads
\begin{equation}
    \Psi^{''} + A_{31}(r) \Psi' + A_{32}(r) \Psi = 0,
\end{equation}
where 
\begin{equation}
    A_{31}(r) = 
    A_{11} + A_{22} - \frac{A_{21}'}{A_{21}},
\end{equation}
\begin{equation}
    A_{32}(r) = A_{11}A_{22}- A_{12}A_{21} + A_{22}'  - \frac{A_{21}'A_{22}}{A_{21}},
\end{equation}
the prime indicates the derivative with respect to $r$.  

\subsection{Solving Methods and Boundary Conditions}
The eigenvalue problems associated with the two linear equations can be efficiently solved using a \textit{relaxation} method \citep{1992nrfa.book.....P}.
A concise description of the algorithm has been provided in Paper I and will not be reiterated here.
In our model, we set $r_{1}=0.8$ and $r_{2}= 2.5 $ as the inner and outer boundary, respectively. 
The eigenvalues are significantly influenced by the initial setup of boundary conditions, which depends on the sign of \textcolor{black}{radial wavenumber $k_r$.} 
\textcolor{black}{In Section \ref{HDWs} \& \ref{MHDDws}, we derive the local dispersion relations in planetary atmosphere and disks. These formulas provide insights into the relationship between the group velocity and the phase velocity, so we carefully select appropriate values of $k_r$ based on the group velocity at the boundaries (Section \ref{GV}). }
If the phase velocity is in the same direction as the group velocity, then the sign of $k_r$ is the direction of wave motion. Otherwise, the sign of $k_r$ is opposite to the direction of wave motion. 
But the sign of $k_r$ only helps us understand the properties of the global modes, and it does not actually rule out the possibility of a solution.
In other words, the results are reasonable as long as they satisfy the dispersion relations.

During our calculations, we find two main kinds of global modes with considerable growth rates, $p$-modes and $r$-modes. 
$p$-modes require a outflow boundary conditions at both sides, which means the group velocities $\mathcal{C}_g <0$ at the inner boundary while $\mathcal{C}_g >0$ at the outer boundary. Thus, according to the dispersion relations \eqref{IA cgx} and \eqref{Fast_cgx} \textcolor{black}{derived in Section \ref{GV}}, we should choose $k_r<0$ at the inner boundary and $k_r>0$ at the outer boundary.
For $r$-modes, we expect waves propagate outward from the planet, which means $\mathcal{C}_g >0$ at the inner boundary and $\mathcal{C}_g >0$ at the outer boundary.
But we should still choose $k_r<0$ at the inner boundary
because the equation \eqref{Rossby cgx} \textcolor{black}{in Section \ref{GV}} shows that the sign of $\mathcal{C}_g$ is opposite to the sign of $k_r$. 
More details on how to distinguish the global mode based on the dispersion relations can be seen in Appendix \ref{Appendix_Dispersion}.

%However, through the calculation of the dispersion relation, it becomes clear that the frequency of slow MHD waves closely approaches the Keplerian frequency. As a result, maintaining a consistent $\Omega_{\rm P}$ across all radii becomes unattainable, thus precluding the formation of a global mode.In light of this, we expect that the outer boundary conditions should be the fast MHD waves. This explanation clarifies why \cite{2013ApJ...770...68B} did not identify new magnetosonic modes but rather observed analogous acoustic modes.
% \textcolor{black}{In Appendix \ref{Appendix_Dispersion} we illustrate in detail how the global modes of a linear system can be distinguished based on the dispersion relation.}

Within the context of BLs, the primary parameters consist of the azimuthal wave number $m$, the width of BL $\delta r$, the spin rate of planet $\Omega_0$, the Mach number $\mathcal{M}$ and the magnetic field strength $\lambda$. 
Throughout our calculations, we explored varying resolutions, successfully attaining sound numerical convergence in our solutions.  
By employing numerical iterations with an initial guess solution, the correct eigenvalue $\omega = \omega_r + i \omega_i$ can be determined. 
Here, the real part $\omega_r$ signifies the frequency where the disturbance propagates as a wave, while the imaginary part $\omega_i$ represents the growth rate of the amplitude enhancement over time.

\section{Waves and instabilities}
\label{Sec3}
In the investigation of global modes, adopting a local Cartesian coordinate system proves to be a favorable approximation for examining wave behavior. 
Consequently, curvature terms involving $r^{-1}$ can be neglected, and $k_r$ and $k_{\phi}$ can be replaced by $k_x$ and $k_y$, respectively.
Assuming suitable initial conditions and performing algebraic manipulations, we can attain approximate dispersion relations for the planet and the disk.

\subsection{Hydrodynamic Waves in Planetary Atmosphere}\label{HDWs}
\textcolor{black}{The magnetic pressure is much lower than the exponentially increasing gas pressure in planetary atmospheres. Thus, we can still utilize the dispersion relation derived in Paper I for the case of pure fluids.}
Specifically, under isothermal conditions ($\gamma=1$), the equation can be written as
\begin{equation}\label{inner dispersion}
\begin{split}
    \sigma^3 - \left[ c_s^2\left(k_x^2+k_y^2 \right) + \frac{c_s^2}{4 H^2}+ 4\Omega^2 \right] \sigma + 2g \Omega k_y = 0 \ , \
\end{split}
\end{equation}
where $H = RT/g = c_s^2/(\gamma g)$ is the scale height of atmosphere, $T$ is the temperature and $R$ is the universal gas constant.
While obtaining specific solutions requires solving a cubic equation, simple approximations can provide insight into the characteristics of these solutions.
In the high-frequency regime, the last term can be neglected, leading to the dispersion relation for inertial-acoustic waves
\begin{equation}\label{iadip}
\sigma^2_{\rm IA} = \left[ c_s^2\left(k_x^2+k_y^2 \right) + \frac{c_s^2}{4 H^2}+ 4\Omega^2 \right] \ . \
\end{equation}
Noting that in the case $\Omega=0$ and $H \rightarrow \infty$, the above equation reduces into the dispersion relation of acoustic waves, namely 
\begin{equation}
    \sigma^2 = (k_x^2 + k_y^2) c_s^2 \ . \
\end{equation}
Hence equation \eqref{iadip} describes two branches of acoustic waves propagating in opposite directions under the influence of rotational effects and vertical stratification.
On the other hand, in the low-frequency regime, the first term can be neglected and we have the dispersion relation for Rossby waves as
\begin{equation}
    \sigma_{\rm R} = \frac{ (2/\gamma) H \Omega k_y }{H^2\left(k_x^2+k_y^2 \right) + \frac{1}{4}+ (4\Omega^2 H^2/c_s^2) }  \  .  \
\end{equation}
This is a new kind of wave introduced in the presence of the planetary rotation, which vanishes at $\Omega = 0$.

\subsection{MHD Density Waves in Disks}\label{MHDDws}
In the context of a pure fluid, perturbations propagate as density waves throughout the disk. 
However, the introduction of magnetic fields complicates the problem, giving rise to various types of MHD waves. 
These waves are essentially distinct from hydrodynamic waves, even in the presence of weak magnetic fields.
Now we consider a rigidly rotating disk
\footnote{
\textcolor{black}{
The exact dispersion relation under differential rotation is complicated, and rigid rotation is a good approximation when we simply analyze the local properties of waves. 
Appendix \ref{Appendix_Dispersion} shows that this approximate dispersion relation matches well with the calculated results.
}
}
with uniform density, pressure, and magnetic field. Using the WKB approximation, we obtain the following dispersion relation
(Appendix \ref{AppendixA})
\begin{align}
     (\sigma^2 - k_y^2 c_a^2)
& \left\{
\sigma^4
-\left[
(c_s^2 + c_a^2)(k_x^2 + k_y^2)+\kappa^2 
\right]\sigma^2
\right. \nonumber \\
& \quad  \left. 
+ (k_x^2 + k_y^2)k_y^2 c_a^2 c_s^2
\right\}
= 0 \ . \ 
\end{align}
The former corresponds to Alfv{\'e}n waves with 
\begin{equation}
    \sigma^2_{\rm A} = k_y^2 c_a^2 \ . \
\end{equation}
The latter corresponds to slow and fast magneto-acoustic waves. The specific dispersion relation can be written as 
\begin{align}\label{sfws}
    \sigma^2_{\rm S,F} &= \frac{1}{2}
\left[
(k_x^2 +k_y^2)(c_s^2 + c_a^2)
+\kappa^2
\right]
 \nonumber \\
& \quad  
\left[
1
\pm
\left(
1
- 
\frac{4 (k_x^2 + k_y^2)k_y^2 c_a^2 c_s^2}
{\left[
(k_x^2 +k_y^2)(c_s^2 + c_a^2)
+\kappa^2
\right]^2}
\right)^{1/2}
\right] \ , \ 
\end{align}
where the branch with the `$-$' sign refers to the slow modes and the branch with the `$+$' sign refers to the fast modes.
For a weak magnetic field, the term 
\begin{equation}
    \frac{4 (k_x^2 + k_y^2)k_y^2 c_a^2 c_s^2}
{\left[
(k_x^2 +k_y^2)(c_s^2 + c_a^2)
+\kappa^2
\right]^2}
\ll 1,
\end{equation}
and we could get approximate expressions for two modes. One corresponds to the slow MHD density waves, given by
\begin{equation}\label{SMHD}
    \sigma^2_{\rm S} = 
 \frac{(k_x^2 + k_y^2) k_y^2 c_s^2c_a^2}
{(k_x^2 + k_y^2)(c_s^2 + c_a^2)
+\kappa^2} .
\end{equation}
Another one represents the fast MHD density waves, written as
\begin{equation}\label{FMHD}
    \sigma^2_{\rm F} = (k_x^2 +k_y^2)(c_s^2 + c_a^2)
+\kappa^2 .
\end{equation}
We can see that the above two dispersion relations are identical to the approximate formulas derived in \cite{1998ApJ...493..102L} when self-gravity is neglected.

\subsection{Group Velocities}\label{GV}
The group velocity along the x-direction can be defined as $\mathcal{C}_g = \partial \omega/\partial k_x = \partial \sigma/\partial k_x$. Substituting equation \eqref{iadip} into it, we obtain the expression for inertial-acoustic waves
\begin{equation}\label{IA cgx}
    \mathcal{C}_g^{\rm IA} =  
    \frac{ k_x c_s^2 }{ \left[c_s^2(k_x^2+k_y^2)+\frac{c_s^2}{4H^2} + 4 \Omega^2 \right]^{1/2}} .
\end{equation}
Similarly, for Rossby waves we have
\begin{equation}\label{Rossby cgx}
    \mathcal{C}_g^{\rm R} = - \frac{(4 c_s^2 g \Omega k_y) k_x }{\left[c_s^2(k_x^2+k_y^2)+\frac{c_s^2}{4H^2} + 4 \Omega^2 \right]^2} .
\end{equation}
Exact $\mathcal{C}_g$ of slow and fast MHD waves can be acquired from equation \eqref{sfws}. 
When starting from dispersion relations \eqref{SMHD} and \eqref{FMHD}, we have approximate expressions
\begin{equation}\label{Slow_cgx}
    \mathcal{C}_{g}^{\rm S} = \frac{k_x  k_y^2 c_s^2c_a^2 \kappa^2 }
{(\omega - m\Omega) \left[(k_x^2 + k_y^2)(c_s^2 + c_a^2)
+\kappa^2\right]^2},
\end{equation}
and 
\begin{equation}\label{Fast_cgx}
    \mathcal{C}_g^{\rm F} = \frac{k_x(c_s^2+c_a^2)}{\omega - m\Omega} .
\end{equation}

\subsection{Resonances}
When these equations are reformulated using the Lagrange displacement $\boldsymbol{\xi}$, a series of singularities emerge in the coefficients of the second-order wave equation. The positions of these singularities serve as indicators of $\textit{resonances}$ \citep{2015ApJ...802...54U,2003MNRAS.341.1157T}.
One can identify these singularities by examining the coefficients in the equations \eqref{leq1} and \eqref{leq2}.

\textit{Corotation resonance} (CR) arises when $\sigma = 0$, indicating a match between the rotational speed and the pattern speed $\Omega = \Omega_{\rm P}$. 
Analysis of the angular velocity profile reveals the presence of two CRs.
The inner corotation resonance (ICR) resides within the BL, approximately at $r = 1$, while the outer corotation resonance (OCR) is located at  
\begin{equation}
     r_{\rm CR} = \Omega_{\rm P}^{-2/3}
\end{equation}

$\textit{Lindblad resonance}$ (LR) occurs when the local frequency matches the radial epicyclic frequency, as described by
\begin{equation}
    m^2(\Omega - \Omega_{\rm P})^2 = \kappa^2 .
\end{equation}
The positions of these resonances in disks can be determined by 
\begin{equation}\label{LR}
    r_{\rm LR}=\left ( \frac{m}{m \pm 1} \Omega_{\rm P} \right )^{-2/3} ,
\end{equation}
Each CR is accompanied by two LRs on either side, referred to as the inner Lindblad resonance (ILR) and the outer Lindblad resonance (OLR).
Although these definitions may not be applicable in the presence of magnetic fields, they can serve as a good approximation for weakly magnetic fields.

Two additional resonances are introduced by magnetic fields.
The first one, known as the \textit{magnetic resonance} (MR), corresponds to the local frequency matching the frequency of slow MHD waves
\begin{equation}
    m^2(\Omega - \Omega_{\rm P})^2 = \frac{k_{\phi}^2 c_a^2 c_s^2}{c_a^2+c_s^2}.
\end{equation}
This equation shares the same form as equation (19) in \cite{2015ApJ...802...54U}, which represents the scenario in the absence of a vertical field. 
Similarly, another resonance emerges only when there is a match between the local frequency and the frequency of Alfv{\'e}n waves 
\begin{equation}
    m^2(\Omega - \Omega_{\rm P})^2 = k_{\phi}^2 c_a^2 ,
\end{equation}
which is referred to as the \textit{Alfv{\'e}n resonance} (AR). 
However, it should be emphasized that this occurs exclusively when $k_z \neq 0$.

In the limit where the disk aspect ratio  $h/r \ll 1$, the positions of the MR and the AR can be determined by first-order approximation. They are
\begin{equation}
    r_{\rm MR} = r_{\rm CR}  \pm \frac{2 h}{3 \sqrt{1 + \beta}},
\end{equation}
and 
\begin{equation}
    r_{\rm AR} = r_{\rm CR}  \pm \frac{2 h}{3 \sqrt{\beta}},
\end{equation}
where $h$ and $\beta$ can be estimated in the vicinity of $r_{\rm CR}$
\citep{2015ApJ...802...54U,2003MNRAS.341.1157T}.
Obviously, their locations are independent of $m$ but rely on the shape of the disk and the intensity of magnetic fields.
The MR lies inside the AR, and both resonances converge to the CR as $\beta \rightarrow \infty$.

\subsection{Turning Points and Propagating Properties}\label{TPaPP}
\begin{figure}[t]
    \centering
    \includegraphics[width=1\linewidth]{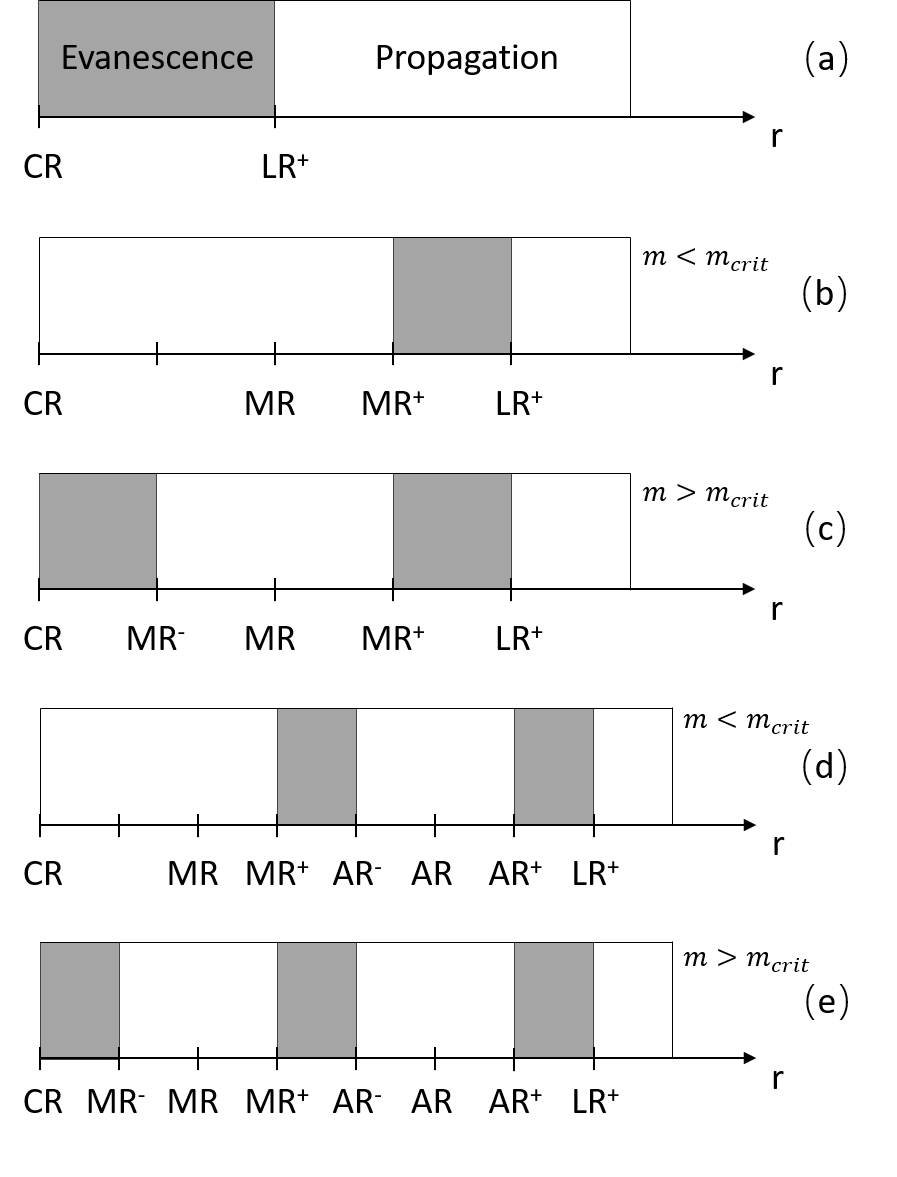}
    \caption{Wave propagation under different conditions.
    Panels (a)(b)(c) correspond to 2D disks $(k_z= 0)$, while panels (d)(e) correspond to 3D disks $(k_z \neq 0)$. 
    Only pure fluids are considered in panel (a), while the other cases involve magnetic fields.}
    \label{WP}
\end{figure}

\textit{Turning points }indicate the transition from wave behavior to non-wave behavior.
By locating these points and resonances, we can identify the propagable and evanescent regions of waves. 
The precise information regarding the position distribution under the influence of an azimuthal magnetic field is provided by \cite{2015ApJ...802...54U} through analytical methods.
In Figure \ref{WP}, we present a simplified schematic diagram to illustrate these positions. Panels (a)(b)(c) correspond to 2D disks with $k_z= 0$, while panels (d)(e) correspond to 3D disks with $k_z \neq 0$. 
It is worth mentioning that case (a) is solely concerned with pure fluids, whereas the remaining cases incorporate the presence of magnetic fields. 
Waves are capable of propagating in white areas, while they become evanescent in shaded areas.

In panel (a), it is evident that there is only a single turning point, known as the effective LR.
The positions of the effective LR can be influenced by magnetic fields, and they also deviate from the nominal LR at high values of $m$. 
Here we consistently denote these points as $\rm LR^+$. 
Between the CR and the LR, all the radial wavenumber become purely imaginary, resulting in the inability of waves to propagate.

In panels (b) and (c), where magnetic fields are introduced, additional turning points emerge on both sides of the MR, labeled by the $\rm MR^-$ and the $\rm MR^+$.
Of particular interest is that the $\rm MR^-$ exists only when $m$ exceeds a critical value, $m_{\rm crit}$. 
Numerical results based on toroidal magnetic fields give a dependency relation $m_{crit} \simeq \sqrt{\beta} \times r/h$
\citep{2008ApJ...679..813M}
, implying a change in the propagation properties at higher $m$ or larger $\lambda$. 
In the section \ref{AMF_Profile}, we will delve into their differences in the transport of angular momentum. 
Under the condition of $m < m_{\rm crit}$, there is only one evanescent region $\rm (MR^+, LR^+)$. However, a new evanescent region emerges near the CR when $m > m_{\rm crit}$, bounded by the $\rm MR^-$.

Similarly, in panels (d) and (e), the presence of the AR introduces two turning points, identified as the $\rm AR^-$ and the $\rm AR^+$.
It is worth noting that the $m_{\rm crit}$ here is also dependent on $k_z$, increasing with higher values of $k_z$.
When $m < m_{crit}$, there are two evanescent regions, including $\rm (AR^+, LR^+)$ and $\rm (MR^+, AR^-)$.
However, when $m > m_{crit}$, the region inside $\rm MR^-$ also becomes evanescent once again. 

\subsection{Axisymmetric Stability}
To investigate the local stability under axisymmetric ($m=0$) perturbations, \cite{2009ApJ...702...75Y} provided a sufficient condition known as the generalized Rayleigh criterion
\begin{equation}
    \kappa^2 (r) + \Omega_{a,r}^2 \ge 0,
\end{equation}  
where
\begin{equation}
    \Omega^2_{a,r} =
    k_{\phi}^2c_a^2 +
    c_a^2 
    \left(
    \frac{d \ln \rho }{d r} + \frac{2}{r}
    \right)
    \left(
    \frac{c_s^2 }{c_a^2 +c_s^2 }\frac{d \ln \rho }{d r}
    \right)
\end{equation}
for the case of $\alpha =1$.
Evidently, this term remains positive irrespective of the monotonicity of the density profile, whether increasing or decreasing. 
Moreover, a stronger magnetic field results in a larger value of $\Omega^2_{a,r}$, implying the stabilizing effect of magnetic fields.
Another term involving rotation, $\kappa^2 \geq 0$ holds for all regions, so that the sum of the two terms is always greater than 0, thereby confirming the stability of our initial setup against axisymmetric perturbations.

\subsection{Non-axisymmetric Instabilities}
In the case of non-axisymmetric ($m \neq 0$) perturbations, the conservation of angular momentum for individual fluid elements is no longer valid due to the force induced by the azimuthal gradient ($\partial / \partial \phi$). 
As a result, the shear term ($\kappa^2$) no longer stabilizes the system but instead promotes instabilities.

\subsubsection{Rossby Wave Instability}
For a thin, non-magnetized Keplerian disk, \citet{1999ApJ...513..805L} argued that the initial radial profile is unstable to non-axisymmetric waves if the function
\begin{equation}
    \mathscr{L}(r) \approx \frac{\Omega}{\kappa^2}\Sigma^{2/\Gamma-1}T^{2/\Gamma}
\end{equation}
attains an extreme value within the domain. 
This phenomenon, known as the Rossby wave instability (RWI),
can lead to the emergence of Rossby vortices and the formation of spiral shocks. 

The criterion can be simplified to $\mathscr{L}(r) \approx \Omega\Sigma/\kappa^2$ in the isothermal process with $\Gamma = 1$ and constant $T$. 
Based on the initial setup outlined in Section \ref{ISet}, the presence of a minimum within the BL can be clearly demonstrated.
Subsequent numerical simulations further validate the presence of the RWI in the non-linear regime, where they delve into various properties, including morphological characteristics and angular momentum transport
\citep{2022MNRAS.509..440C,2022MNRAS.512.2945C}.
\textcolor{black}{As mentioned in \cite{2009ApJ...702...75Y}, the growth of the RWI could be suppressed in the presence of magnetic fields. For example, the $m = 5$ outer edge mode gradually weakens with increasing magnetic field and vanish at $\lambda \sim 0.9$. 
% Nevertheless, for our problem under weak magnetic fields $\lambda= 0 \sim 0.4$, the RWI is not significantly suppressed.
}

It is worth mentioning that the occurrence of vortices in compressible disks can also be attributed to the Papaloizou-Pringle instability  \citep{1984MNRAS.208..721P}. However, this instability necessitates a power-law index  $q$ in rotation profile $\Omega(r) \sim r^{-q}$ that exceeds $\sqrt{3}$, rendering it invalid for Keplerian rotation ($q=3/2$).

\subsubsection{Magnetorotational Instability }
In a 3D homogeneous disk with weakly magnetic fields, the occurrence of the magnetorotational instability (MRI) requires that the angular velocity profile satisfies the condition
\begin{equation}
    \frac{d \Omega^2}{d r} < 0,
\end{equation}
\citep{1998RvMP...70....1B}. 
However, given that the gradient of angular velocity within the planet is $d\Omega /dr = 0$ and within the BL is $d\Omega /dr >0$, the MRI cannot work in these regions. However, the MRI is applicable to the Keplerian disk where $\Omega \propto r^{-3/2}$.

In the investigation of a Keplerian disk featuring a gap and subjected to azimuthal magnetic fields, the work conducted by \cite{2009ApJ...702...75Y} revealed the MRI in the vicinity of the disk boundaries. 
The growth rate of the MRI exhibits a positive correlation with $c_s$ and reaches its maximum at a moderate $\lambda$ or $m$. However, these unstable modes are suppressed and eventually vanish when the parameter $\beta$ is low. 
Remarkably, they found that this phenomenon only occurs when using rigid boundary conditions but remains absent under outflow boundary conditions or $k_z=0$. 
In our calculations, we also cannot find the MRI using the outflow boundary conditions.

Numerical simulations also reveal that global modes emerge at early times, while the MRI develops effectively within the disk at later times \citep{2013ApJ...770...68B}. This phenomenon may imply that the growth rate of MRI in the linear phase is small, or that the strength of the magnetic field capable of forming MRI has not yet fully developed.
Based on the above analysis, we infer that the global modes identified in Paper I will continue to dominate during the linear phase, albeit with some local properties being modified by the presence of the magnetic field. 
However, the MRI will also gradually develop due to increasingly intensified magnetic fields.

\section{Boundary Layer Modes}\label{BLMoSP}
Local perturbations can propagate across the entire system, thereby giving rise to global modes that exhibit a consistent frequency $\Omega_{\rm P}$.
In Paper I, we have identified two classes of global modes, namely the inertial-acoustic modes ($p$-modes) and the Rossby modes ($r$-modes)
\footnote{\color{black}{Although the RWI and $r$-modes share the same name, Rossby, they are not related to each other. The $r$-modes only comprise the Rossby waves in the planetary atmosphere and the density waves in the disk. However, the $p$-modes comprise the inertial-acoustic waves in the planetary atmosphere, the RWI within the BL and the density waves in the disk.
When the planet is non-rotating, $r$-modes vanish while $p$-modes survive, and the inertial-acoustic waves become acoustic waves.}}.
The disparity between these two types of modes is reflected in the characteristics of internal waves. 
One involves sound waves influenced by rotational effects, where the group velocity aligns with the phase velocity. 
However, Rossby waves have the group velocity opposite to the phase velocity. 
% For the sake of simplicity, we shall term them as $p$ and $r$ modes.

\textcolor{black}{In this section, we would study $p$-modes in the presence of an azimuthal magnetic field. 
The basic parameters of $p$-modes are $m= 15, \delta r = 0.05, \Omega_0 = 0, \mathcal{M} =6, \lambda = 0.2, k_z =20 $, and results are $\Omega_{\rm P} = 0.799$ and $\rm Im[\omega] = 0.173
$.
We are not going to describe the r-modes too much here. Since we find that $r$-modes have similar wave behaviors in the BL and the disk compared to $p$-modes, and the wave behavior in planetary atmospheres is almost unchanged from that presented in Paper I. 
This is because the gas pressure grows exponentially, much higher than the magnetic pressure, and therefore the wave behaviour is more hydrodynamic.
However, the magnetic field may play a role in the BL and the disk where the gas pressure is relatively not very high.
}

\subsection{Perturbation Quantities and Spatial Morphology}
\label{PQaSM}
\begin{figure*}[t]
\begin{minipage}{0.5\textwidth}
    \centering
    % \subfigure[]
    {
    \centering
    \includegraphics[width=1\linewidth]{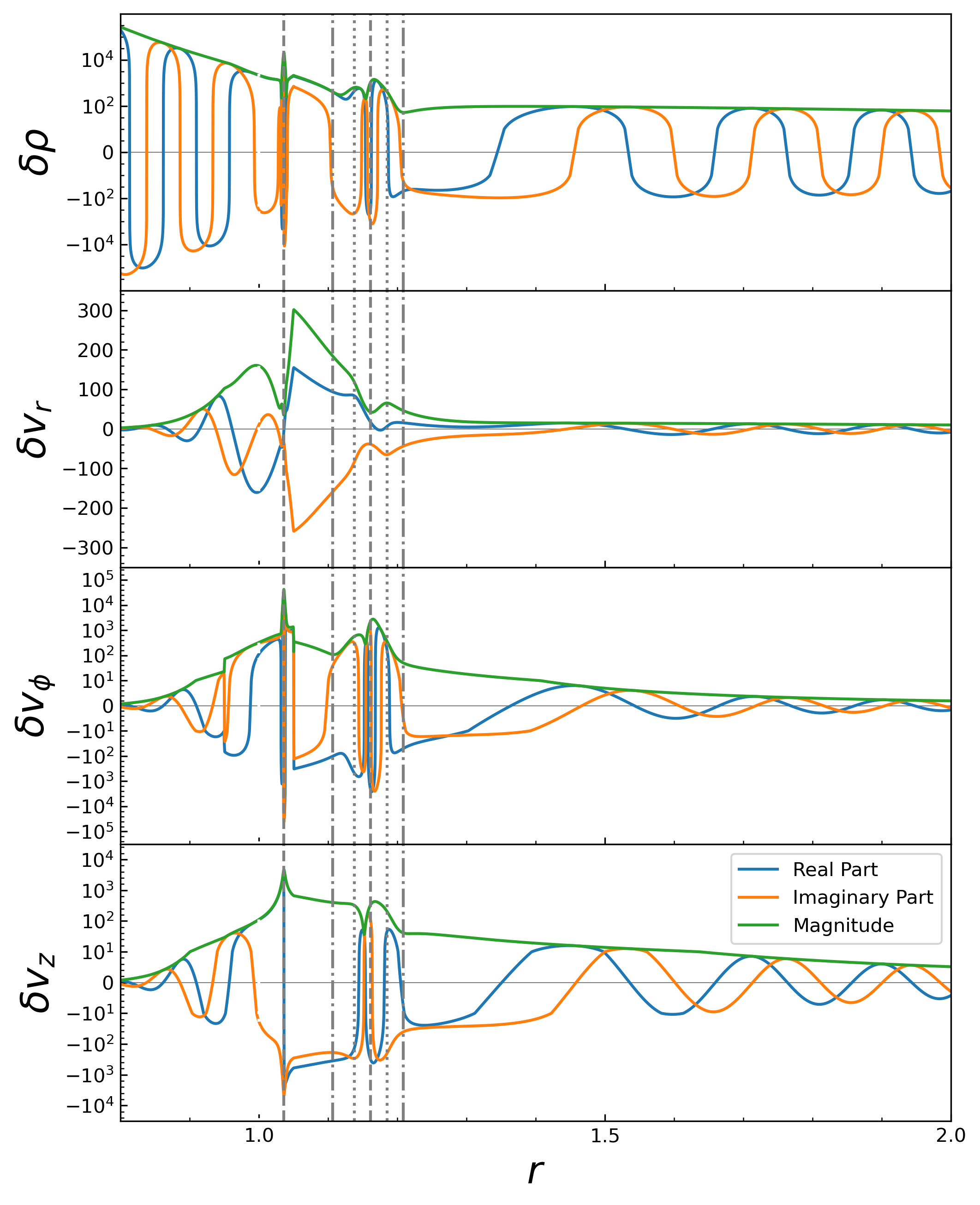}
    }
\end{minipage}
\begin{minipage}{0.5\textwidth}
     % \subfigure[]
    {    
    \centering
    \includegraphics[width=1\linewidth]{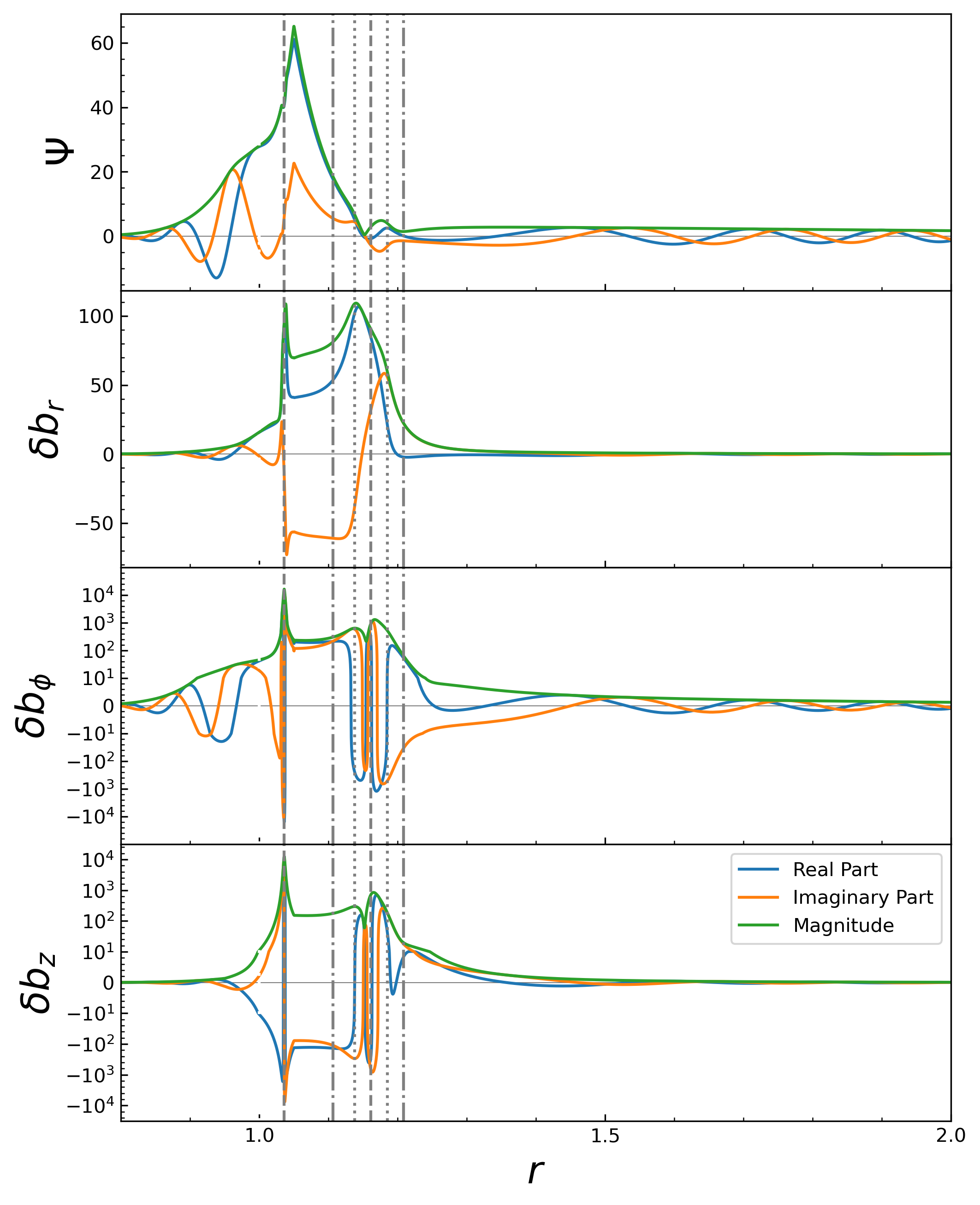}
    }
\end{minipage}
    \caption{Radial distribution of different perturbations corresponding to the $p$-mode. The dashed lines correspond to the CR, the dotted lines correspond to the MR and the dash-dot lines represent the LR.}
    \label{perturbation-profile}
\end{figure*} 
When compared to the profiles of pure fluids presented in Paper I, the influence caused by the magnetic field primarily manifests near the CR. To illustrate this, we plot the radial profile of perturbations in Figure \ref{perturbation-profile}.
These profiles encompass perturbations of density $\delta \Sigma$, pressure relative to density $\delta \Psi$, velocity $\delta v_r$, and magnetic field $\delta b_r$ across three spatial directions $(r,\phi,z)$.
In this figure, the real component is represented by blue lines, the imaginary component by orange lines, and the magnitude by green lines.
The dashed lines correspond to the CR, while the dotted lines correspond to the MR. 
The dash-dot lines represent the LR, which are calculated using an approximate formula for pure fluids. 
The illustration of the AR is omitted here due to its close proximity to the MR.
% The magnetic disturbance lacks significance for $r<1.05$, as the established magnetic field is confined exclusively to the disk.

\textcolor{black}{Recall the overview and discussion of perturbations under pure fluids in Paper I.
We can see that the amplitude of the real or imaginary part of the perturbation oscillates sinusoidally in the planetary atmosphere ($r<1$) or in the disk ($r>1.5$), implying that the perturbation propagates as waves.
However, in the evanescent regions between resonances, the amplitude of the real and imaginary parts no longer show sinusoidal variations and therefore no longer behave as waves.
Between the CR and the LR, there is a smooth transition from sinusoidal to non-sinusoidal curves, which also marks a shift from wave-like behavior to non-wave behavior.
However, results under magnetic fields in this paper show a reappearance of sinusoidal curves, suggesting that this region no longer remains evanescent.
}
% Between the CR and the LR, the transition from sinusoidal to non-sinusoidal curves occurs smoothly, which also signifies a shift from wave-like behavior to non-wave behavior.
These phenomena can be elucidated through our analysis in Section \ref{TPaPP}. Panel (a) of Figure \ref{WP} corresponds to the case of pure fluids, there is only one evanescent region located between the CR and the LR, hence the smooth curve transition.  Conversely, the remaining panels demonstrate that the influence of magnetic resonance can segregate the region into alternating sequences of evanescent and propagative regions so that wave behavior reappears.

Of significance is the observation that the perturbations $\delta \rho$, $\delta \Psi$, and $\delta v$ attain their maximal amplitudes in the vicinity of the ICR. This occurrence suggests the emergence of vortices near the BL, a phenomenon linked to the RWI.
Interestingly, an additional peak emerges near the MR with a relatively weaker amplitude. This region marks the maximum perturbation in the magnetic field. Consequently, the magnetic field will be excited and progressively intensified in this area as time progresses, thereby leading to a change in the original distribution of magnetic fields.

\begin{figure}[t]
    \centering
    \includegraphics[width=1\linewidth]{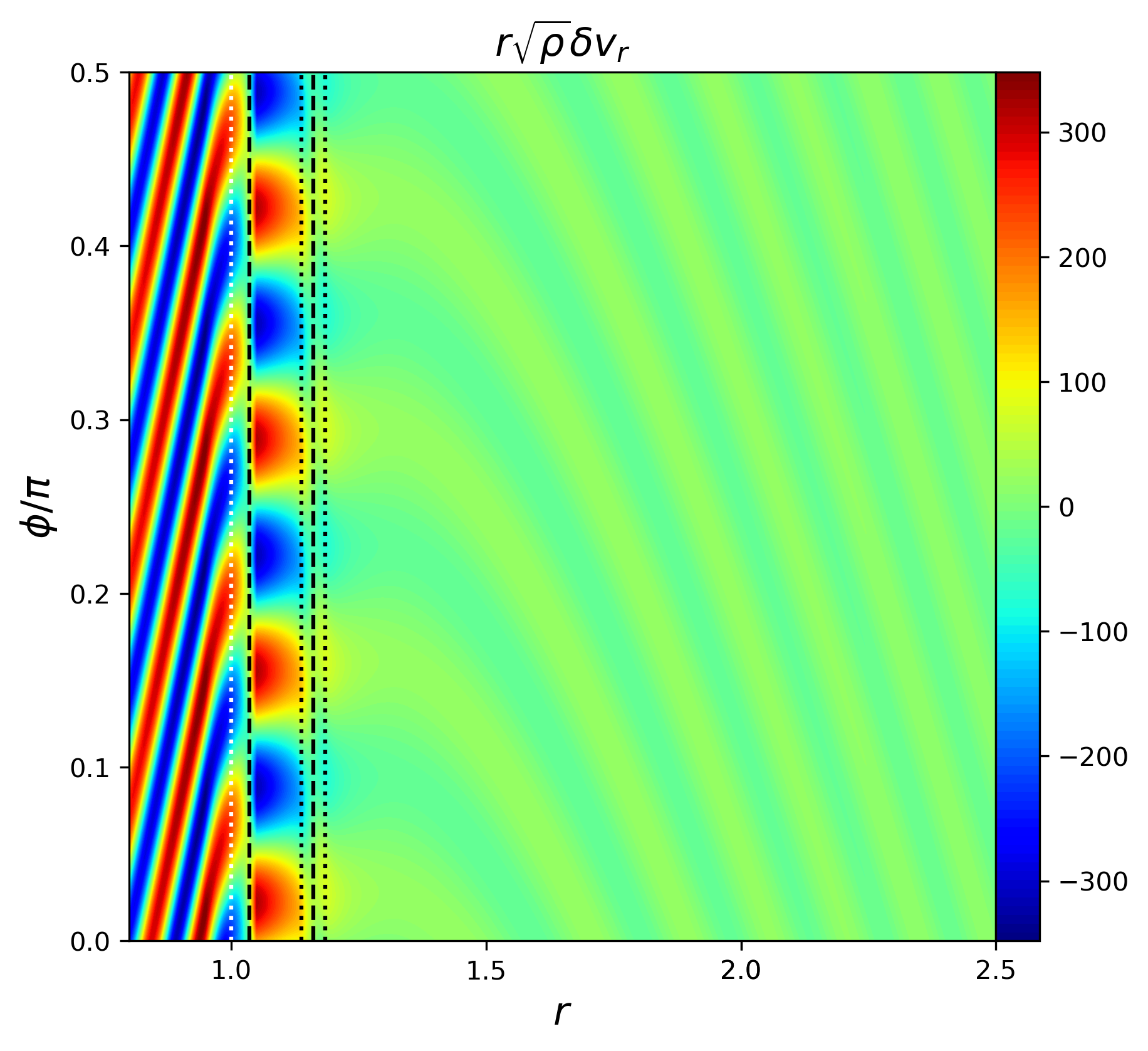}
    \caption{Spatial morphology of the $p$-mode illustrated by quantities $r\sqrt{\rho}\delta v_r$.  The white dotted line indicates the surface of a planet, the black dashed lines correspond to the CR and the black dotted lines correspond to the MR.
    }
    \label{Radiation Pattern}
\end{figure}

\textcolor{black}{The quantity $r\sqrt{\rho}\delta v_r$ can be used in the visualization of wave patterns,
as it exhibits conservation during the propagation of waves in planets and disks. 
In Figure \ref{Radiation Pattern} we show the snapshots of $r\sqrt{\rho}\delta v_r$ in the $r-\varphi$ coordinate plane.
As we can see, this image bears a prominent resemblance to Figure 9 presented in \cite{2013ApJ...770...67B}. 
}
% Compared with the snapshots illustrated in Paper I, the $r$-modes have displayed negligible alterations, so we just show the structure of $p$-mode (corresponding to parameters D1 of Table \ref{AMF_para}) in Figure \ref{Radiation Pattern}. 
Broadly, it can be divided into three parts, including  
Inclined waves within planets, spiral density waves within disks and vortices near the BL.
It is noted that these vortices are confined between two CRs due to the influence of magnetic resonances. 
By contrast, under the scenario of pure fluids, the spatial confinement of vortices transcends the OCR, extending as far as the OLR.

\subsection{Angular Momentum Transport}\label{secAMF}
\begin{figure*}[t]
    \centering
    \includegraphics[width=1\linewidth]{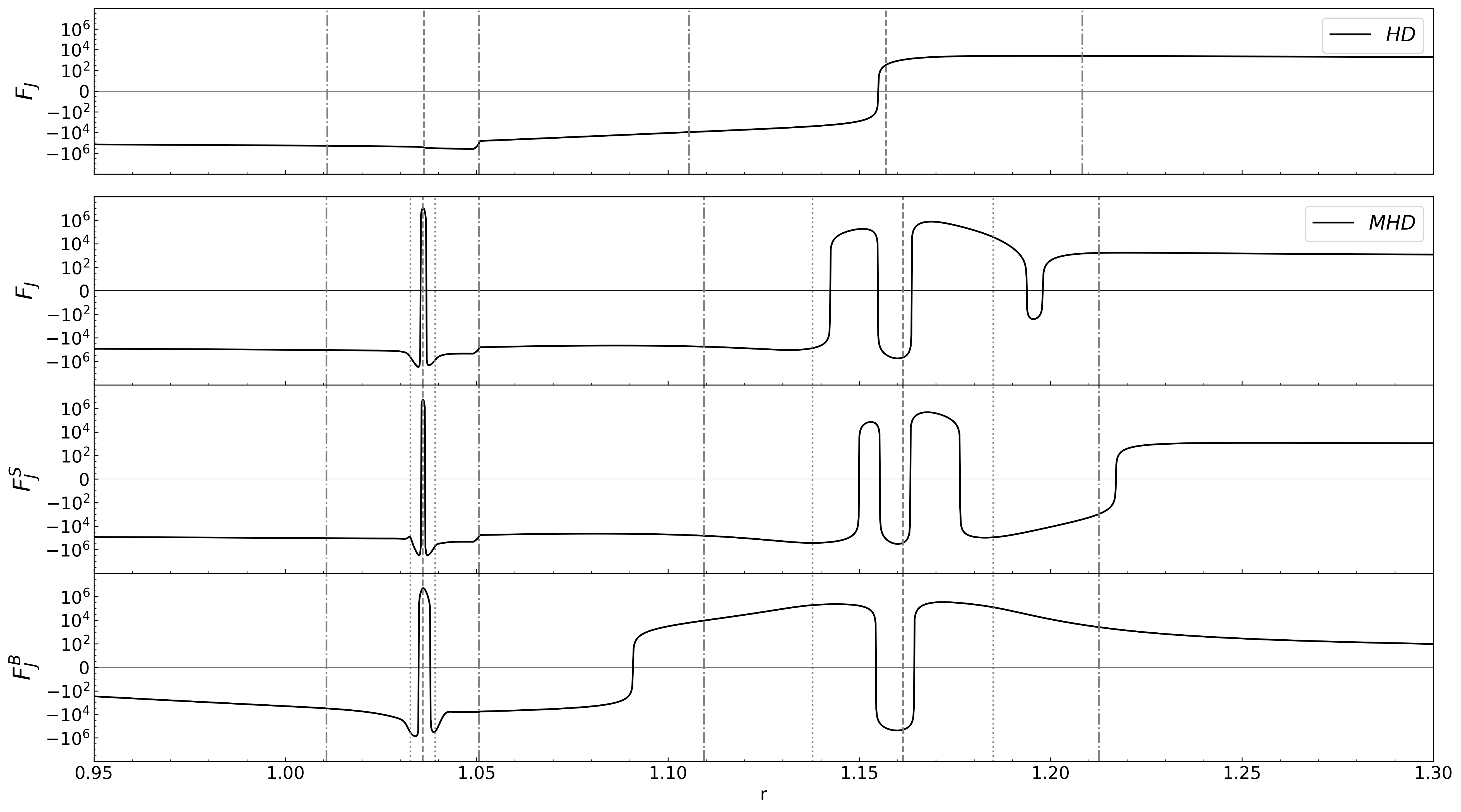}
    \caption{Distribution of the AMF $F_J$ over distance $r$ for $p$-modes.
    The top two panels show the total AMF for the HD and MHD cases,  respectively, and the third and forth panels show the stress and magnetic terms for the MHD case. The dashed lines correspond to the CR, the dotted lines correspond to the MR, and the dashdotted lines correspond to the LR. 
    }
    \label{AMF}
\end{figure*}
In Paper I, we have illustrated the distribution of angular momentum flux (AMF) across the entire region. 
Inside the planet, the AMF is negative for $p$-modes while positive for $r$-modes.
Consequently, $p$-modes would spin up the planet while $r$-modes would slow down the planet.  
In other regions, both modes exhibit analogous patterns, characterized by a positive AMF in the disk and a negative AMF situated between the disk and the planet.
In our subsequent analysis, we will take $p$-modes as examples to investigate the impact of the magnetic field on the AMF.

Broadly speaking, the net AMF of a system consists of three components: advective, stress, and magnetic terms. This decomposition can be written as
\begin{equation}
    \mathcal{F}_{\rm J} =
    \mathcal{F}_{\rm J}^{\rm A}
    + 
    \mathcal{F}_{\rm J} ^{\rm S} + 
    \mathcal{F}_{\rm J}^{\rm B}
\end{equation}
\citep{1999MNRAS.307..645F}. The advective term is expressed as 
\begin{equation}
     \mathcal{F}_{\rm J}^{\rm A} = 
     r^2 \Sigma \Omega
     \int_{0}^{2\pi} d\theta 
     \text{Re}[\delta v_r],
\end{equation}
where $\text{Re}[...]$ denotes the real part of physical quantities.
When considering the first order of perturbations with $\delta f \sim e^{im\phi}$, the integration yields 0 as a result of the periodicity in the azimuthal angle. Therefore, we can ignore the advective term in the case of waves. 
For the second order of perturbations, the stress term can be written as 
\begin{equation}
    \mathcal{F}_{\rm J} ^{\rm S} = r^2 \Sigma \int_{0}^{2\pi} d\theta \text{Re}[ \delta v_r]  \text{Re}[\delta v_{\theta}] \ , \
\end{equation}
which can be explicitly formulated in terms of perturbation quantities \citep{2002ApJ...565.1257T,2008gady.book.....B} as
\begin{equation}
     \mathcal{F}_{\rm J} ^{\rm S} = \pi r^2 \Sigma
\left(
\text{Re}[\delta v_r ]\text{Re}[\delta v_{\theta}]
+ \text{Im}[\delta v_r ] \text{Im}[\delta v_{\theta}]
\right) \ . \
\end{equation}
Similarly, we can establish an analogous relationship for the AMF through magnetic fields, namely
\begin{equation}
    \mathcal{F}_{\rm J}^{\rm B} = - r^2 \int_{0}^{2\pi} d\theta 
    \text{Re}[\delta b_r] 
    \text{Re}[\delta {b}_{\theta}] \ , \
\end{equation}
which reduces to 
\begin{equation}
    \mathcal{F}_{\rm J}^{\rm B} = - \pi {r^2}
\left(
\text{Re}[\delta b_r ]\text{Re}[\delta b_{\theta}]
+ \text{Im}[\delta b_r ] \text{Im}[\delta b_{\theta}]
\right) \ . \
\end{equation}
For all the AMF, $\mathcal{F}_{\rm J}<0$ implies the inflow of angular momentum, whereas $\mathcal{F}_{\rm J}>0$ implies the outflow of angular momentum. 

\textcolor{black}{In Figure \ref{AMF}, 
we plot the total angular momentum $\mathcal{F}_{\rm J}$ versus distance for the MHD case (the second panel), and the third and fourth panels show the contribution of the stress and magnetic terms. Meanwhile, we also plot the HD case without magnetic fields as a comparison (the top panel). The dashed lines indicate the CR and the dotted lines indicate the MR.
% In the HD case, the AMF profile varies smoothly with distance.
% However, the distribution of $\mathcal{F}_{\rm J}$ in the vicinity of the CR changes significantly due to the presence of the MR.  Positive $\mathcal{F}_{\rm J}$ appears near the ICR, while multiple alternating positive and negative values occur near OCR.
}

\textcolor{black}{
Overall, the behaviour of the AMF in the presence of a magnetic field is similar to that of a pure fluid, except near the CR.
In the case of a pure fluid, the transition of the AMF from a positive to a negative value occurs only once. However, when the presence of a magnetic field is taken into account, these transitions begin to occur multiple times.
As we discussed in section \ref{TPaPP}, these changes are most likely related to the resonances introduced by the magnetic field.
This change can be a significant impediment to the motion of the fluid elements.
As discussed in \cite{2008ApJ...679..813M}, the strong magnetization of a disk can slow down the migration of planets.
Here, we assume that a group of fluid elements are migrating inward or outward. These fluid elements are unaffected as they pass through the non-magnetized regions, but eventually stall near the magnetic resonances. Finally, the aggregation of gas could form a series of pressure bumps.  
}

\section{Impact of magnetic field strength}\label{Impact_Mag}
In the prior work, we demonstrated that both $p$ and $r$ modes exist in a rotating planet-disk system. 
The dominant modes determine whether the planet spins up or spins down, as well as whether accretion or release of gas. 
As various parameters can affect the development of global modes, we will delve into how the growth rates of these two types of modes vary with the magnetic field strength $\lambda$ in this section.

\subsection{Growth Rates}
\begin{figure*}[htbp]
    \centering
    \begin{minipage}[t]{0.48\textwidth}
        \centering
            \subfigure[$p$-modes]
        {
        \centering
        \includegraphics[width=1\linewidth]{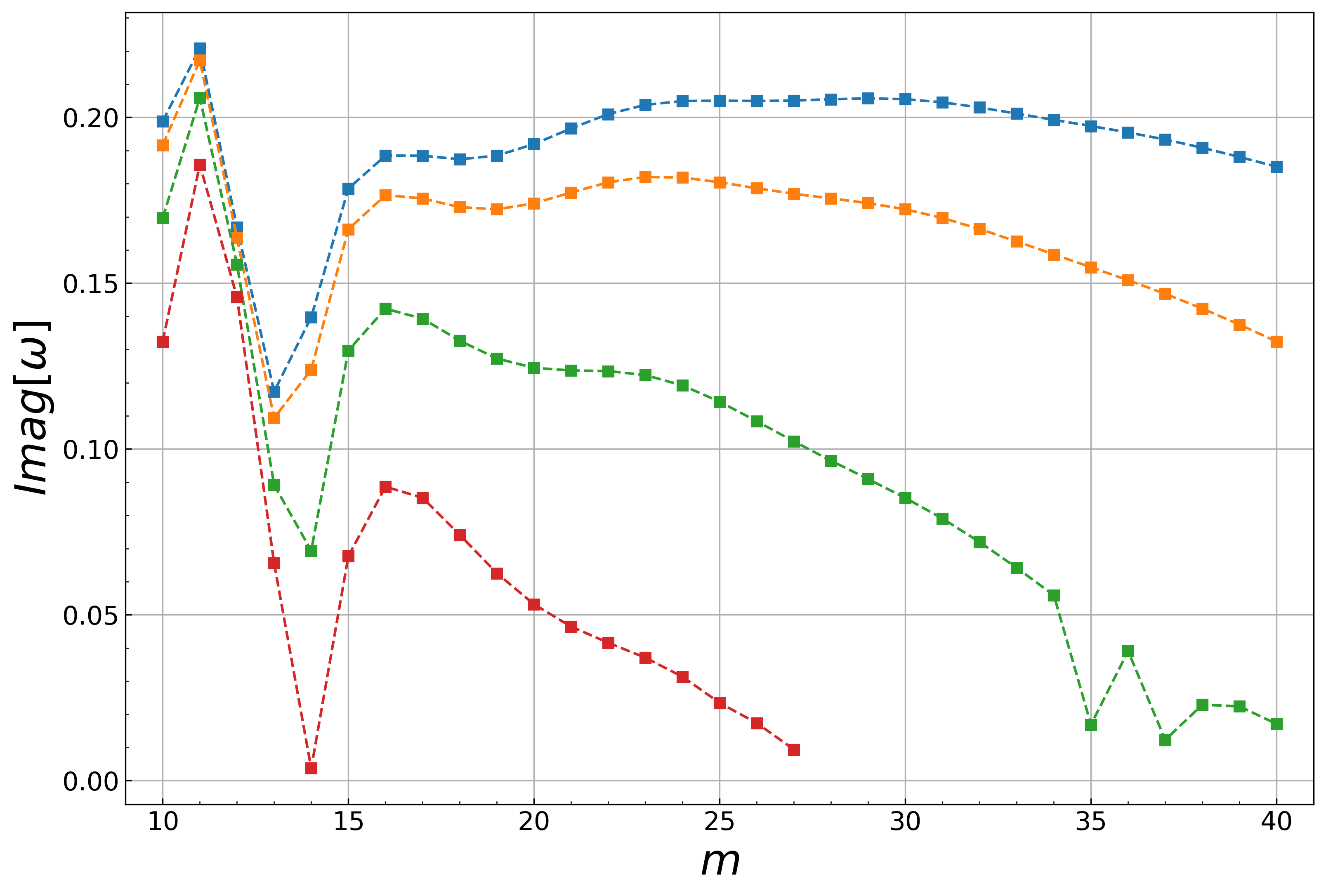}
          }
    \end{minipage}
    \hfill
    \begin{minipage}[t]{0.48\textwidth}
        \centering
            \subfigure[$r$-modes]
        {
        \centering
        \includegraphics[width=1\linewidth]{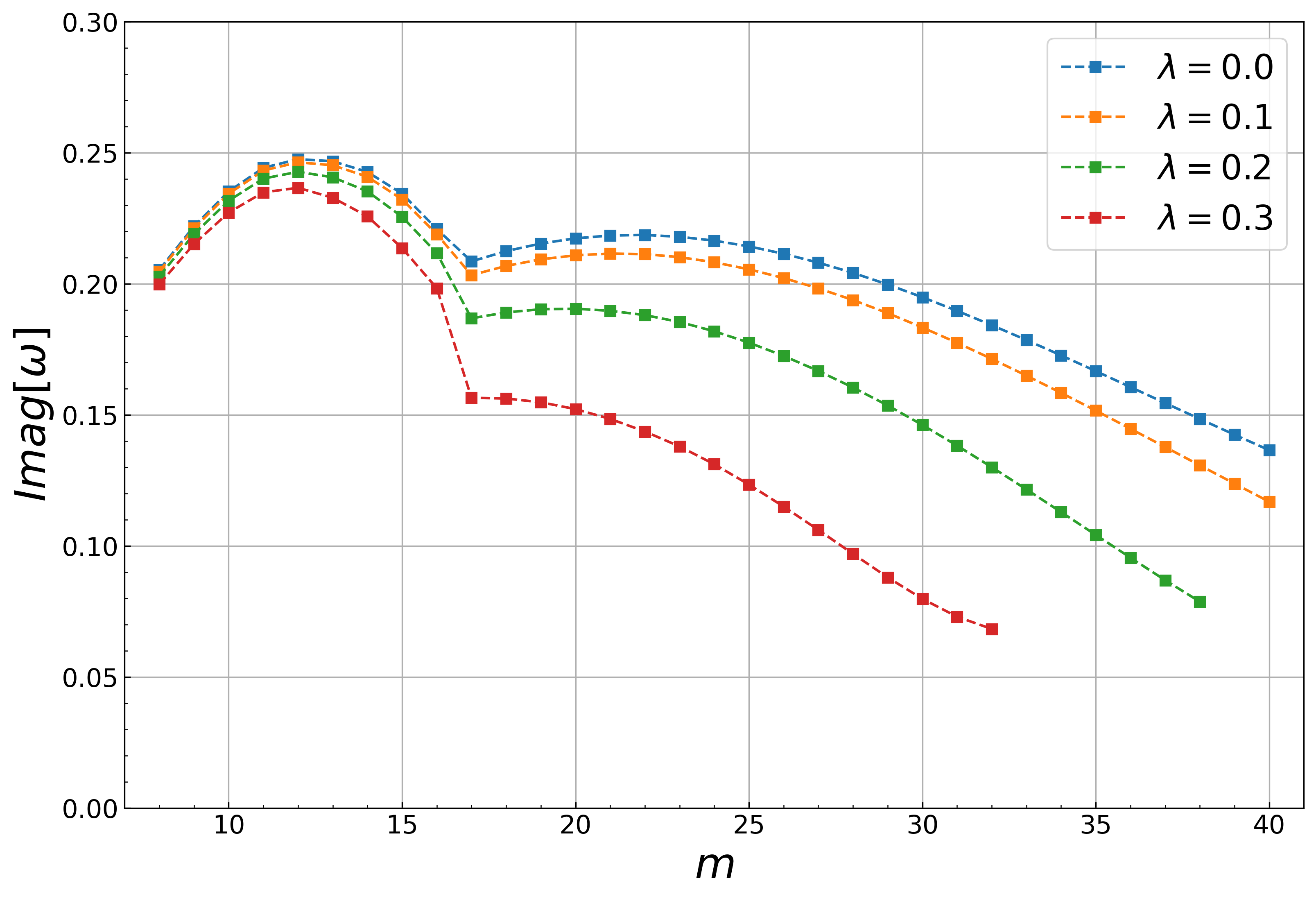}
          }
    \end{minipage}
    \caption{Dependence of growth rates $\rm {Imag}[\omega] $ on the azimuthal mode number $m$ under different magnetic field strength $\lambda$. The left panel shows the results of $p$-modes and the right panel shows the results of $r$-modes.
    }
    \label{Modes-mag}
\end{figure*}

\begin{figure*}[htbp]
    \centering
    \begin{minipage}[t]{0.48\textwidth}
        \centering
            \subfigure[$m=11$]
        {
        \centering
        \includegraphics[width=\linewidth]{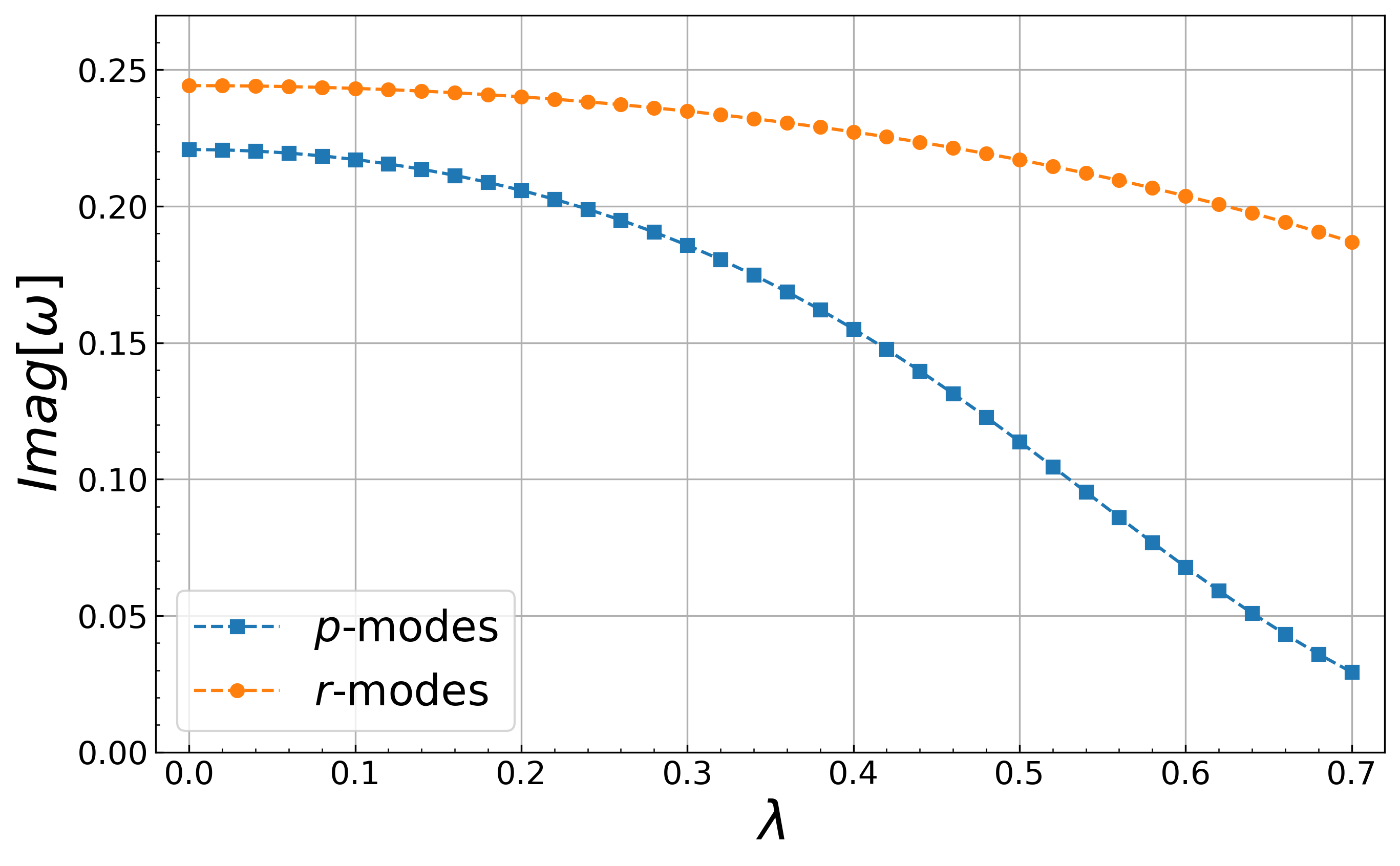}
          }
        \label{subfig:c}
    \end{minipage}
    \hfill
    \begin{minipage}[t]{0.48\textwidth}
        \centering
            \subfigure[$m=25$]
        {
        \centering
        \includegraphics[width=\linewidth]{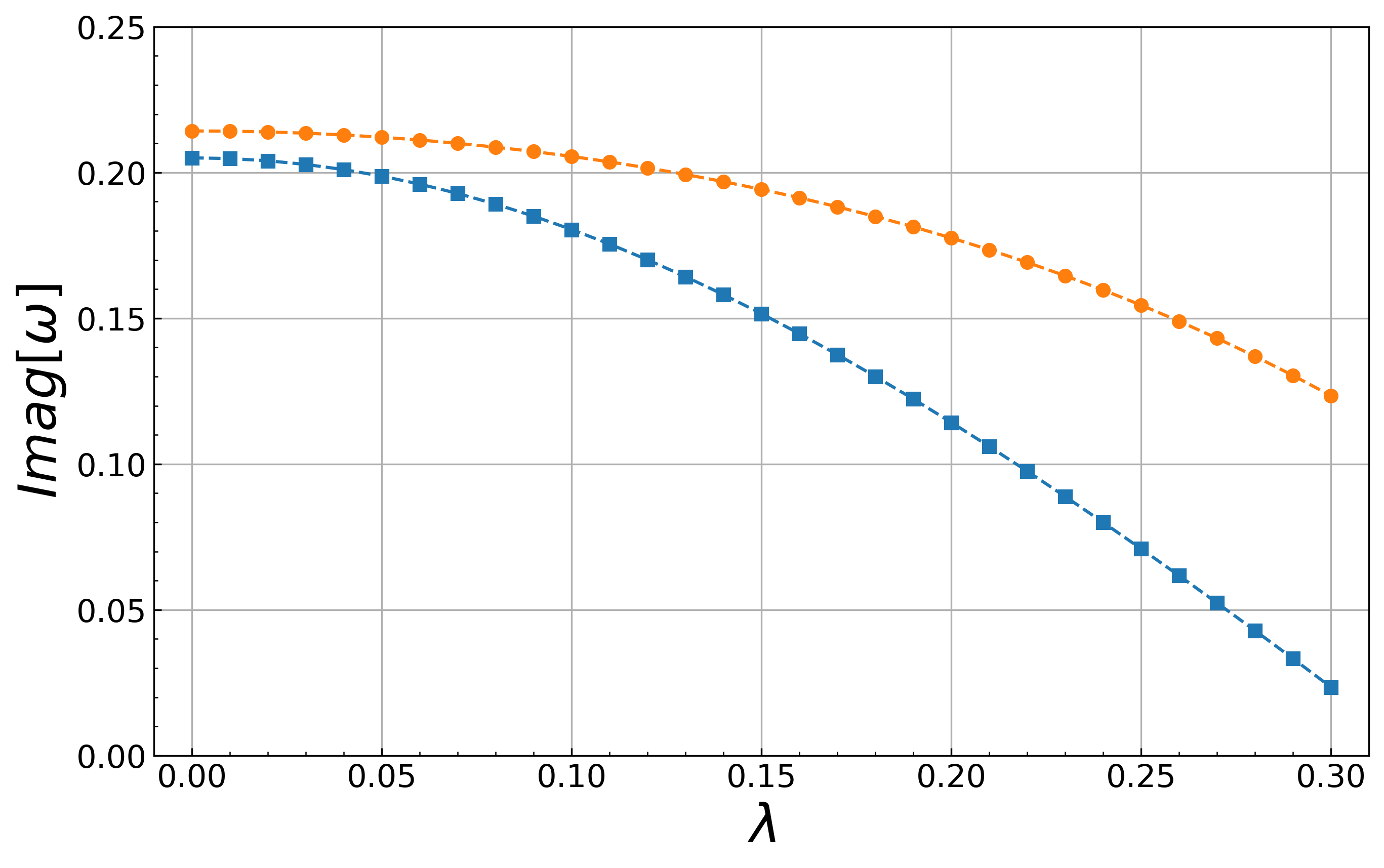}
          }
        \label{subfig:d}
    \end{minipage}
    \caption{Dependence of growth rates Im$[\omega]$ on the magnetic field strength $\lambda$.}
    \label{Growth Rate-mag}
\end{figure*}

Figure \ref{Modes-mag} shows the dependence of growth rates on various azimuthal mode number $m$ under $\lambda = \{0.0, 0.1, 0.2, 0.3\}$. The basic parameters used in these two modes are $\delta r=0.05$, $\mathcal{M}=6$ and $\Omega_0=0.5$. 
Over a range of tens of values of $k_z$, we do not find significant changes in the growth rate. This is consistent with the fact that the dominate modes concentrated at $k_z = 0$, as mentioned by \cite{2013ApJ...770...67B}.
Hence, we maintain $k_z = 0$ in studying dependence of growth rates. 
For cases of $\lambda = 0.0$, we can see that the growth rates are mainly distributed in two regions, around $m=11$ and $m=25$. 
As $\lambda$ increases, the growth rate of larger $m$ decreases more significantly than that of lower $m$.

For modes of $m=11$ and $m=25$, we further show how they change with the magnetic field strength $\lambda$ in Figure \ref{Growth Rate-mag}. The trend of decreasing growth rates with increasing $\lambda$ is consistent for both $p$ and $r$ modes. However, the attenuation of $r$-modes is weaker than $p$-modes. 
This is because the atmosphere plays a major role in $r$-modes, where the gas pressure is much stronger than the magnetic pressure, so $r$-modes are less affected by magnetic fields. On the other hand, $p$-modes are more susceptible to the magnetic fields because they originate from the boundary layer where the gas pressure is not very high.

It is worth noting that $p$-modes are strongly suppressed at around $\lambda \sim 0.3$ for $m=25$, but the value can be up to $\lambda \sim 0.7$ for $m=11$. This phenomenon is similar to the study of growth rates of the RWI demonstrated by \cite{2009ApJ...702...75Y}. The $m=5$ outer edge mode of the RWI is strongly suppressed at around $\lambda \sim 0.88$, implying that there may be a close relationship between the $p$-mode and the RWI.

\subsection{Transition Points between Spin-up and Spin-down}
\begin{figure}[htbp]
    \centering
    \includegraphics[width=1\linewidth]{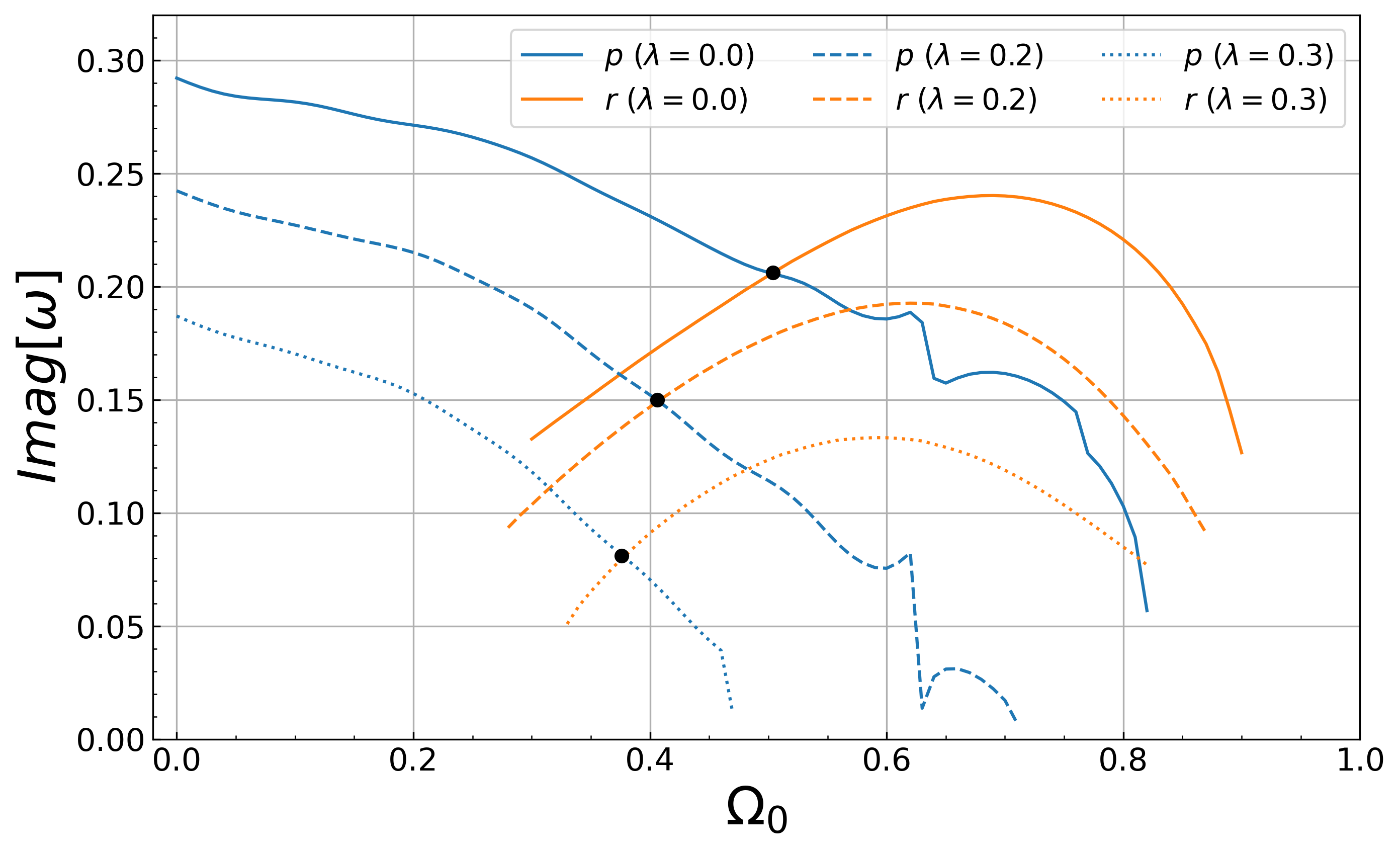}
    \caption{Dependence of growth rates $\rm {Imag}[\omega] $ on the planetary spin $\Omega_0$ under different magnetic field strength $\lambda$. The blue lines represent $p$-modes while the orange lines represent $r$-modes. The black dots indicate the turning points between $p$-mode dominance and $r$-mode dominance.}
    \label{Omega-grw}
\end{figure}

% In a previous study, we showed that the planetary spin $\Omega_0$ can affect the growth rate of both modes.
We now study the effect of the magnetic field strength $\lambda$ on the global modes of various spin rates $\Omega_0$, as shown in Figure \ref{Omega-grw}. The parameters used in these cases are $m=25$, $\delta r=0.05$, $\mathcal{M}=6$ and $k_z = 0$.
At low $\Omega_0$, the $p$-mode dominates, so the planet gains angular momentum and rotates faster. On the contrary, the $r$-mode dominates at high $\Omega_0$, so the planet loses angular momentum and rotates slower.
As $\lambda$ increases from 0.0 to 0.2 and 0.3, we can see a decrease in the growth rate for all modes.  
In particular, the transition points $\Omega_t$ that mark the dominance of the $p$ mode and the $r$ mode move to lower values. From $\Omega_t \sim 0.5$ at $\lambda =0.0$ to $\Omega_t \sim 0.41$ at $\lambda =0.2$ and $\Omega_t \sim 0.38$ at $\lambda =0.3$.

\subsection{AMF Profiles}\label{AMF_Profile}

\begin{figure}[htbp]
    \centering
    \includegraphics[width=1\linewidth]{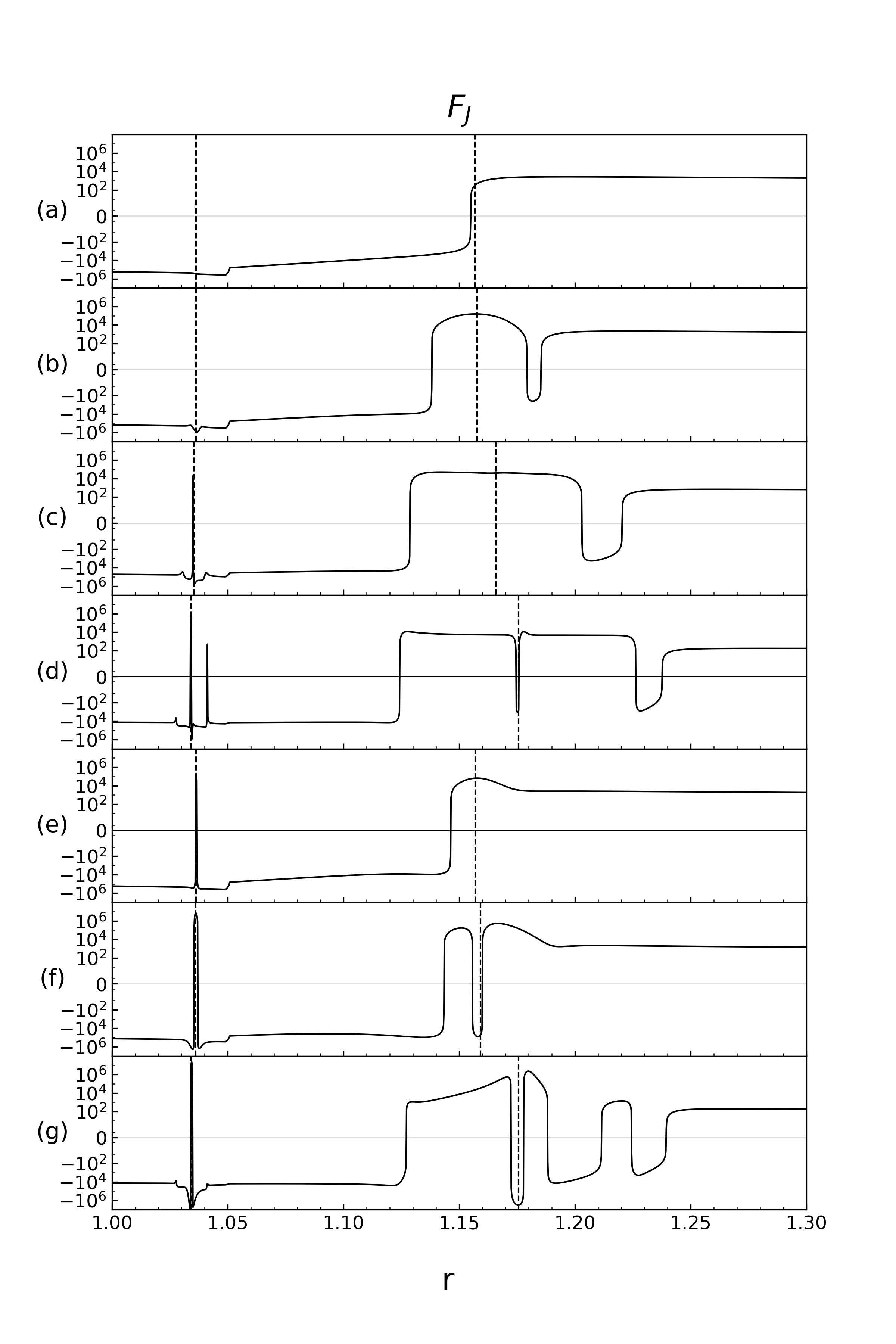}
    \caption{Distribution of the AMF $F_J$ in the vicinity of the CR for $p$-modes. The dashed lines correspond to the CR. Panel (a) shows the HD calculations. Panels (b)-(d) are the 2D MHD calculations at $\lambda = \{0.09, 0.3, 0.45 \}$  while panels (e)-(g) are the 3D MHD calculations at $\lambda = \{0.03, 0.15, 0.45 \}$. }
    \label{AMF_total}
\end{figure}

Next, we focus on the variation of the AMF profile near the CR, as shown in Figure \ref{AMF_total}. The basic parameters are $m=15$, $\delta r = 0.05, \mathcal{M}=6$ and $\Omega_0 =0$.
Panels (a)-(d) show the 2D calculations, where 
panel (a) shows the HD results as comparison, while panels (b)-(d) show the MHD results at $\lambda = \{ 0.09, 0.3, 0.45 \}$.
Panels (e)-(f) correspond to 3D MHD results ($k_z = 20$) at $\lambda = \{0.03, 0.15, 0.45\} $.

For 2D cases, the first change occur near the OCR when $\lambda > 0.09 $ (panel (b)). When $\lambda >0.3$, the second change occur near the ICR.
As for 3D cases, the first change occur near the ICR when $\lambda > 0.03$ (panel (e)), whereas the second change appear near the OCR when $\lambda> 0.15$ (panel (f)).
Compared to the negative $F_J$ shown in panel (a), these changes near the ICR are the appearance of positive $F_J$. In the vicinity of the OCR, the changes are transitions between positive and negative $F_J$.
As we mentioned in section \ref{TPaPP}, the MR would emerge near the CR if we take into account the magnetic field, and the AR would also emerge if $k_z \neq 0 $. These resonances and their turning points can form a series of evanescent and propagable zones, which explains the variations of $F_J$ near the CR.
Notice that there are more turning points when $m > m_{crit}$. Since $m_{crit} \simeq \sqrt{\beta} \times r/h$, these profiles change further as $\lambda$ increases (panels (d) and (g)).

\section{Discussion}\label{Sec5}

\textcolor{black}{Previous sections illustrate global modes under magnetic fields in several aspects including morphology, angular momentum transport and growth rates.
Although in this paper we mainly focus on problems of planets, our model is applicable to other systems as long as they are consistent with planet-disk structure with the BL (i.e. the central object can be a white dwarf, a neutron star or even a minor planet). 
We now discuss the connection between theory and numerical simulations as well as observations for different astrophysical systems.}
\subsection{Implications: Amplification of Magnetic Fields near the BL}
\cite{1989MNRAS.236..107P} propose that the presence of strong shear in the BL can induce the development of toroidal magnetic fields. 
Using 3D MHD numerical simulations, \cite{2002MNRAS.330..895A} confirm the amplification of magnetic fields in boundary layer problems involving a non-rotating, unmagnetized star.
Furthermore, \cite{2013ApJ...770...68B} investigate the effect of diverse initial field geometries within the disk, including toroidal, vertical, and vertical fields with zero net flux.
They also detect the amplification of the magnetic field in the BL.
In a word, the above simulations reveal prominent peaks in the magnetic field profile near the BL and time-dependent behavior of enhancement, which can be explained by our theory.
As shown in Figure \ref{perturbation-profile}, the perturbation of magnetic fields would reach a maximum near the BL during the linear phase, so the intensity of the perturbation would gradually increase as time evolves.  
At the same time, the growth rate is closely related to the properties of the system, so perturbations can show some kind of temporal dependence due to the dynamic evolution of the BL.
% One plausible explanation arises from the contrast in values of $\beta$, the former has a lower value, $\beta \sim 5-20$, similar to our setup, while the latter has a larger value, $\beta \sim 5000$.

\subsection{Implications: Angular Momentum Belts and Lower Terminal Spin}
% In our previous discussion, we mainly considered a weakly magnetized disk.
% However, in real astrophysical systems, it becomes essential to account for the influence of planetary magnetic fields. 
% The actual distribution of magnetic fields in these cases can be intricate, usually assuming a dipole-like configuration with a radial dependence of $B\sim r^{-3}$. 
% For the sake of simplicity, we persist in utilizing the formula \eqref{imag} as our initial profile, with its coverage extending to the inner boundary (see Appendix \ref{AppendixB}).
% As illustrated in Figure \ref{perturbation-profile2}, the profile of perturbations exhibits no significant deviation from those presented in Figure \ref{perturbation-profile}. 
% It is important to highlight the emergence of a new peak in proximity to the ICR, an occurrence attributed to the presence of MRs and ARs.
% Consequently, the amplification of magnetic fields persists within the BL. 
% This observation is in alignment with the findings of \cite{2002MNRAS.330..895A} but contradicts the conclusions drawn by \cite{2017ApJ...835..238B}.
% One plausible explanation arises from the contrast in values of $\beta$, the former has a lower value, $\beta \sim 5-20$, similar to our setup, while the latter has a larger value, $\beta \sim 5000$.

It is noting that a prominent positive peak occur within the BL in terms of the AMF (Figure \ref{AMF}). This suggests the potential formation of angular momentum belts around the planet. 
The accumulation of positive angular momentum can exceed the negative angular momentum transferred to the planet, leading to a substantial reduction in the efficiency of wave-mediated angular momentum transport from disks to planets.
\textcolor{black}{In section \ref{AMF_Profile}, we also point out that it is easier to form angular momentum belts near the BL in the 3D MHD case, as the required magnetic field strength $\lambda$ is weaker than that of 2D MHD cases.} 
The formation of angular momentum belts is in agreement with the findings of \cite{2017ApJ...835..238B}, despite their use of a 3D model with initial vertical magnetic fields ($\beta \sim 6200$) situated outside the BL, which could be the result of a gradual increase in magnetic field strength over time throughout the region.

\textcolor{black}{So far, we have demonstrated that the weakening effect of the magnetic field on the transport angular momentum to the planets comes from two aspects: the reduction in growth rates of $p$-modes and the emergence of angular momentum belts near the BL.
On the other hand, the decay of the growth rate of $r$-modes, which causes the outward propagation of angular momentum from the planet, is weaker than that of $p$-modes (Figure \ref{Growth Rate-mag}). 
In addition, the transition points between spin-up and spin down shift towards lower spin rates (Figure \ref{Omega-grw}). Therefore, we infer that the stronger the magnetic field, the lower terminal spin.
}
\subsection{Implications: Formation of Pressure Bumps}
% \textcolor{black}{The gas giants in our solar system all have rings, and even the dwarf planets (Haumea and Quaoar) and centaurs (Chariklo and Chiron) have been found to have rings \citep{2014Natur.508...72B,2015A&A...576A..18O,2017Natur.550..197S,2023Natur.614..239M}. 
% So, is it possible that they share the same mechanism of ring formation?
% One can explain this possibility from the planet-disk model.}
\textcolor{black}{As discussed in Section \ref{PQaSM}, the accretion process is primarily driven by $p$-modes, which exhibit notable features such as the formation of vortices near the BL, and the formation of spiral arms in the disk. 
These phenomena are consistent with those observed in pure fluids \citep{2023ApJ...945..165F, 2022MNRAS.509..440C, 2022MNRAS.512.2945C}, although the vortices in MHD cases are confined by resonances within a smaller region.
The analysis of AMF profiles in Section \ref{secAMF} also suggests that wave motion is hindered by these resonances. As a result, gas may accumulate gradually near the BL, leading to the formation of several pressure bumps. 
Given that dust particles have a tendency to migrate towards pressure maxima \citep{2020apfs.book.....A}, the dust dynamics could result in the emergence of complex substructures \citep{10.1093/mnras/stad2629}.
}
% In Section \ref{secAMF}, we point out that waves would be trapped in several areas due to the existence of MRs or ARs. In terms of the AMF, it takes on multiple alterations near the CR, so gas would gradually accumulate in these places where the AMF is zero, ultimately creating several local pressure maxima.
% \textcolor{black}{
% Since dust particles have a tendency to migrate towards pressure maxima \citep{2020apfs.book.....A}, the formation of substructures is possible if we take into account dust dynamics.}

% Although the later nonlinear evolution of the system is beyond our linear analysis, the structure of vortice due to the RWI remains prominent as it evolves \citep{2022MNRAS.509..440C,2022MNRAS.512.2945C}.
% Regarding to the RWI affected by dust dynamics, \cite{10.1093/mnras/stad2629} identified two types of instabilities.}
% The Type I instability, triggered by sharp pressure bumps, promotes the concentration of dust particles into gas vortices.
% Conversely, the Type II instability, generated by mild bumps, can maintain a structure of rings with additional clumps.
% \textcolor{black}{In addition, nonlinear interactions such as the streaming instability and the gravitational instability can also facilitate the growth of dust to form satellitesimals. 
% As the gas disk expands, the structure of rings may gradually extend over a large area and form primordial rings. 
% Once these materials are beyond the Roche radius, they may continue to grow and aggregate, thereby triggering the inside-out satellite formation \citep{2012Sci...338.1196C, 2021ApJ...923...27H}. 
% }

\subsection{Implications: Spin-up and Spin-down in Spin Evolution}
Observations of the spin evolution in accreting neutron stars reveal transitions between spin-up and spin-down phases \citep{1997ApJ...481L.101C,2010int..workE..16G}. 
Several theories have been put forth to account for these phenomena. 
For example, accretion is considered a plausible mechanism for events leading to spin-up \citep{2008MNRAS.389..608F}, whereas spin-down might be due to gravitational waves or magnetic braking \citep{1996MNRAS.280..458A,2023MNRAS.518.4322Y}.
\cite{2020MNRAS.492..762W} point out two potential scenarios in the analysis of wind-accreting magnetars, either accretion with outflows or accretion while spinning down. 
Our theory supports the viability of this concept but from the perspective of wave modes.
Because of the coexistence of the $p$-modes and $r$-modes for a rotating central object, the two modes can be considered as different manifestations of accretion. 
Whether spin-up or spin-down depends on the dominant modes.
When $p$-modes dominate, the central object tends to undergo long-term spin-up behavior.
On the other hand, when $r$-modes dominate, it results in long-term spin-down tendencies. 
Moreover, during the transitional state between two dominant modes, the system can even demonstrate a periodic alternation between spin-up and spin-down.

\subsection{Correlation with kHz Quasi-Periodic 
Oscillations}
In some binary systems including neutron stars, the observation of X-ray brightness reveals the presence of specific quasi-periodic oscillations (QPOs).
One well-known phenomenon is the occurrence of twin kHz QPOs, spanning a frequency range from 200 Hz to 1300 Hz.
These oscillations comprise a higher-frequency peak referred to as the upper kHz QPO ($\nu_2$) and a lower-frequency peak referred to as the lower kHz QPO ($\nu_1$).
However, observations indicate that the number of peaks can vary. In certain cases, only a single peak is detectable, whereas in other cases, a pair of peaks is observed \citep{2016IJAA....6...82W}.
It is widely accepted that the frequency $\nu_2$ originates from the orbital motion taking place in the innermost disk.
Around neutron stars, the circular motion gives
\begin{equation}\label{freoforb}
    \nu_{orb} = \sqrt{\frac{GM}{4\pi^2 r_{orb}^3}} \approx 1200(Hz)
    \left( \frac{r_{orb}}{15 \rm km}\right)^{-3/2} \left( \frac{M}{1.4 M_{\odot}} \right)^{1/2}.
\end{equation}
This equation gives the same order of magnitude as the observational data. 

Similar to the interpretations presented by \cite{2003ApJ...591..354T}, it is proposed that Rossby gravity waves can be responsible for generating both lower and upper QPO frequencies.
We suggest that the twin kHz QPOs originate from $p$-modes and $r$-modes. 
\textcolor{black}{In the context of dimensional model, we can take $M= 1.4 M_{\odot} $ and $r_{orb} =  15 r \ \rm km$, so the equation \eqref{freoforb} yields $\nu_{orb} \approx 1200 (Hz) r^{-3/2}$.}
Since the most significant disturbance occurs near the CR, we use the location of CRs as a basis for estimating the frequency range.
\textcolor{black}{The CR can be expressed in terms of pattern speed as $r_{CR} = \Omega_{\rm P}^{-2/3}$, thus $\nu_2 \sim 1200 \rm Hz \times \Omega_{\rm P}$.
Regarding $p$-modes, the typical range for $\Omega_{\rm P}$ spans from 0.7 to 0.9, which corresponds to a high-frequency range of $840-1080 \rm Hz$.}
On the other hand, the frequency range for $r$-modes can be arbitrary, owing to its close association with the rotational speed of central objects. This explains the low-frequency range.
The coexistence of both types of modes serves as an explanation for the occurrence of a twin peak in the frequency spectrum.
However, certain conditions can favor the dominance of one mode over the other, leading to the manifestation of a single peak in the frequency spectrum.
The above discussion can also be applied to explain the QPOs of other compact objects, such as the white dwarf, or the black hole. 
For example, some studies also argue the connection between the high-frequency QPOs of black holes and $p$-modes \citep{2015MNRAS.450.2466Y,2013MNRAS.431.3697F}.

\subsection{Future Prospective}
While our analysis has provided intriguing insights related to observational data and numerical simulations, it is important to acknowledge its inherent limitations.
\textcolor{black}{Here we just consider an ideal MHD model which requires a perfectly coupling of the magnetic field to the gas. However, \citet{2014MNRAS.440...89K} state that the standard constant $α$ disc couples to the field via thermal ionisation only within $30\ R_J$. Accretion through most of the mid-plane is forbidden by strong magnetic diffusivity. 
Moreover, in non-ideal MHD problems, factors such as Ohmic, Hall and Ambipolar diffusevities also need to be taken into account.
Accordingly, our analysis works for the accretion process at the inner part of disks. 
}
When we attempt to estimate the terminal spin through the transition points of dominated modes. The value obtained in this manner is specific to a single mode $m$.
A genuine system comprises a cluster of waves with different modes and frequencies, so our linear analysis is insufficient in determining the exact terminal spin.
It is also beyond the scope of linear analysis when it comes to more intricate non-linear behavior like waves and shocks \citep{2012ApJ...752..115B,2023ApJ...952...89M}, merging of vortices \citep{2022MNRAS.509..440C,2022MNRAS.512.2945C}, and effects of dust on the RWI \citep{10.1093/mnras/stad2629}. 
Besides, how does the stratified structure \citep{2016ApJ...817...62P,2021ApJ...921...54D} and cooling effects \citep{2022MNRAS.514.1733H, 2024arXiv240520367D} influence the evolution is also worthy of in-depth study. The impact of radiation pressure can also be important in accretion flows \citep{2022ApJ...934..111A,2001ApJ...553..987B}. 
Exploring circumplanetary disks with BLs offers exciting prospects, and upcoming research should prioritize numerical simulations. The utilization of advanced computational capabilities empowers us to delve deeper into the intricacies of structural and temporal evolution. 

\section{Summary}\label{Sec6}
Starting from the MHD equation, we perform a linear analysis of isothermal planet-disk systems with the BL. 
Specifically, we consider a magnetized disk with azimuthal magnetic fields but it is non-magnetized for the planet and the BL.
We obtain two equations of first-order linear perturbation, which can be solved by the relaxation method with suitable boundary conditions.
From the perspective of a local Cartesian coordinate system, we derive the dispersion relation for planets and disks using the WKB approximation.
Within the planetary interior, perturbations manifest as inertial-acoustic waves and Rossby waves, while in the disk, they manifest as fast or slow MHD density waves.
Additionally, we also calculate the corresponding group velocities.

Our analysis uncovers intricate wave behavior associated with resonances and turning points.
In hydrodynamics, only the CR and the LR would exist in the system.
However, when magnetic fields are considered, the MR emerges, and the AR also comes into play if $k_z \neq 0$. 
These resonances would introduce several turning points that mark the transition from wave behavior to non-wave behavior. 
Consequently, a sequence of evanescent and propagable regions emerges. These regions can trap waves, thereby contributing to the development of the RWI or the MRI. 

\textcolor{black}{Similar effects of magnetic fields are observed on both $p$-modes and $r$-modes, and we use $p$-modes as an example to elucidate this.
The profiles of magnetic perturbations show the emergence of peaks near the BL due to the presence of MRs or ARs. These peaks imply a gradual increase in the strength of the magnetic field with time evolution, i.e. amplification of magnetic fields.
The spatial morphology illustrated by $r \sqrt{\rho} \delta v_r$ unveils the existence of vortices generated by the RWI. The structure are restricted to the region bounded by two CRs, which is smaller in extent compared to the hydrodynamic case.
In terms of the profiles of AMF, we find some changes would occur near the CR in the presence of magnetic fields. Particularly, positive values of $F_J$ occur near the BL imply the possibility of formation of angular momentum belts around the planet, which can suppress the angular momentum transferred to the planet. Due to the presence of MRs and ARs, it is also possible for the materials to accumulate near the CR and then form a series of pressure bumps.
}

\textcolor{black}{Next, we turn our attention to the impact of magnetic field strength $\lambda$ on the growth rates, transition points between spin-up and spin-down, and the AMF profiles. 
Growth rates of all $m$ are diminished as $\lambda$ increases. 
The lower modes of $m$ are less affected by the magnetic field than the higher modes of $m$, and require a stronger $\lambda$ to be suppressed.
Compared to $p$-modes, $r$-modes are less affected by the magnetic field, and require a stronger $\lambda$ to be suppressed.
Regarding to the effect of planetary spin $\Omega_0$. The $p$-mode dominates at low $\Omega_0$, while the $r$-mode dominates at high $\Omega_0$. 
The dominance of two kinds of modes may be related to the spin evolution of accreting neutron stars with spin-up and spin-down.
After additional consideration of the magnetic field, the turning points between $p$-modes dominance and $r$-modes dominance shift towards a lower values of $\Omega_0$.
In terms of the AMF profiles, we find that the angular momentum belt near the BL can form at lower values of $\lambda$ in 3D MHD cases, while higher values of $\lambda$ are required for the 2D MHD cases. 
Therefore, we infer that the terminal values of equilibrium spin may be lower in the presence of magnetic fields.
Finally, we also discuss the connection between these two modes and twin kHz QPOs, where the $p$-mode may be the upper kHz QPO and the r-mode may be the lower kHz QPO.
}

\textcolor{black}{
Our theory deepens the understanding of the boundary layer accretion in the presence of azimuthal magnetic fields and provides a theoretical basis for observations and numerical simulations.
}
% and subsequently validate the dependency relation expressed as $m_{\rm crit} \simeq \sqrt{\beta}\times r/h$. 
% By adjusting parameters such as $m$ or $\lambda$ to higher values, the AMF profiles can undergo transitions from type B to C or from D to E.
% Then we proceed to examine the impact of the parameters $\delta r$, $\mathcal{M}$, $\lambda$, and $\Omega_0$ on the growth rate.
% Our results show that the magnetic field primarily weakens the $p$-modes while leaving the $r$-modes unaffected. 
% This behavior can be attributed to the distinct origins of these two modes. 
% $p$-modes originate in the disk with a relatively low gas pressure, so the effect of the magnetic pressure is significant.
% On the contrary, $r$-modes originate in the planetary atmosphere characterized by an exponential increase in gas pressure, and therefore the effect of magnetic pressure is negligible.
% As a result, the variations in growth rates closely resemble those observed in the case of pure fluids. Notably, the transition from $p$-modes dominated to $r$-modes dominated shift to a lower value of 0.4, indicating a lower terminal spin for magnetized planets.
\appendix
\section{Dependence of Global Modes on Resolution}\label{Resolution}
\begin{figure*}[t]
    \centering
    \begin{minipage}[t]{0.48\textwidth}
        \centering
            \subfigure[$p$-modes]
        {
        \centering
        \includegraphics[width=\linewidth]{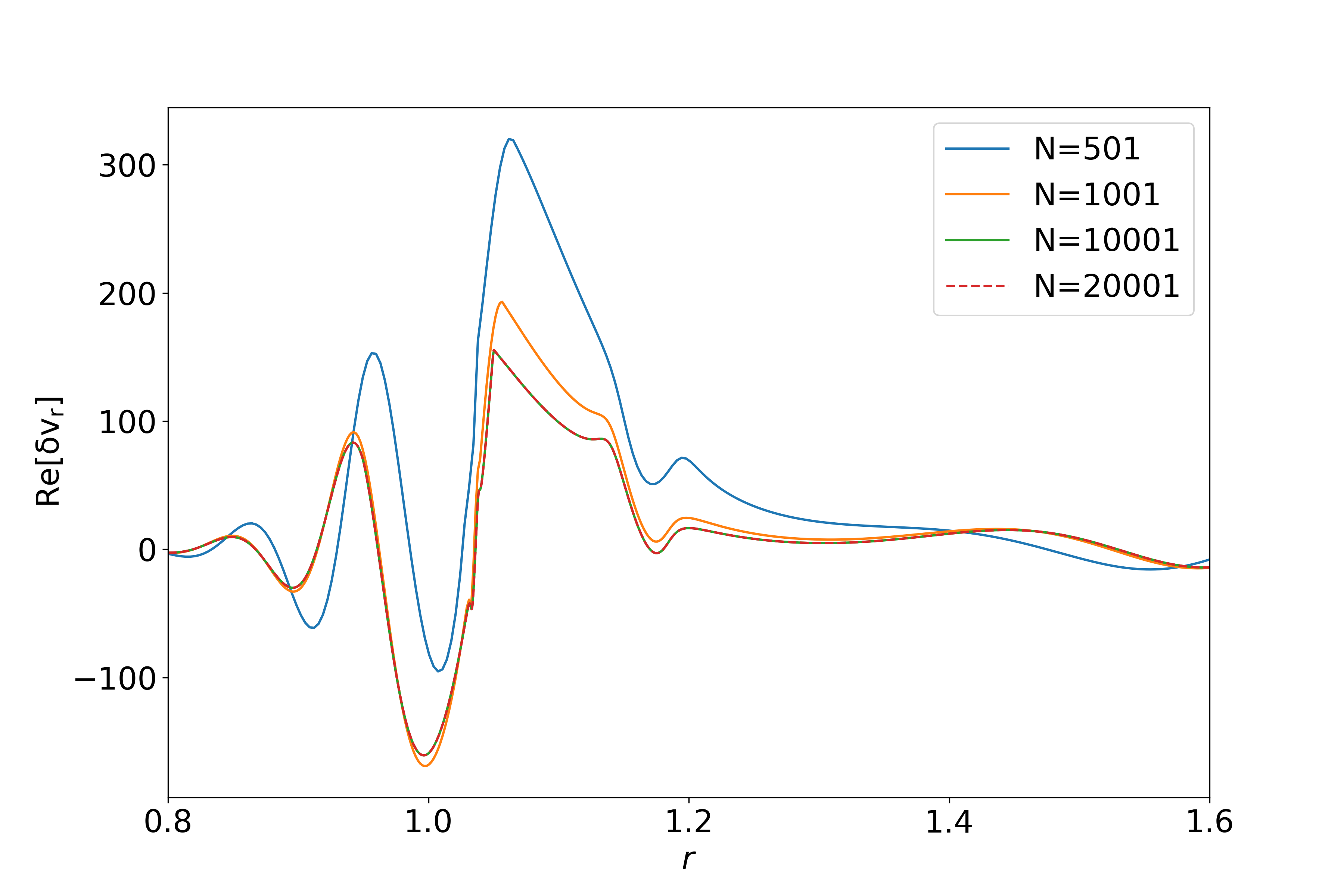}
          }
        \label{}
    \end{minipage}
    \hfill
    \begin{minipage}[t]{0.48\textwidth}
        \centering
            \subfigure[$r$-modes]
        {
        \centering
        \includegraphics[width=\linewidth]{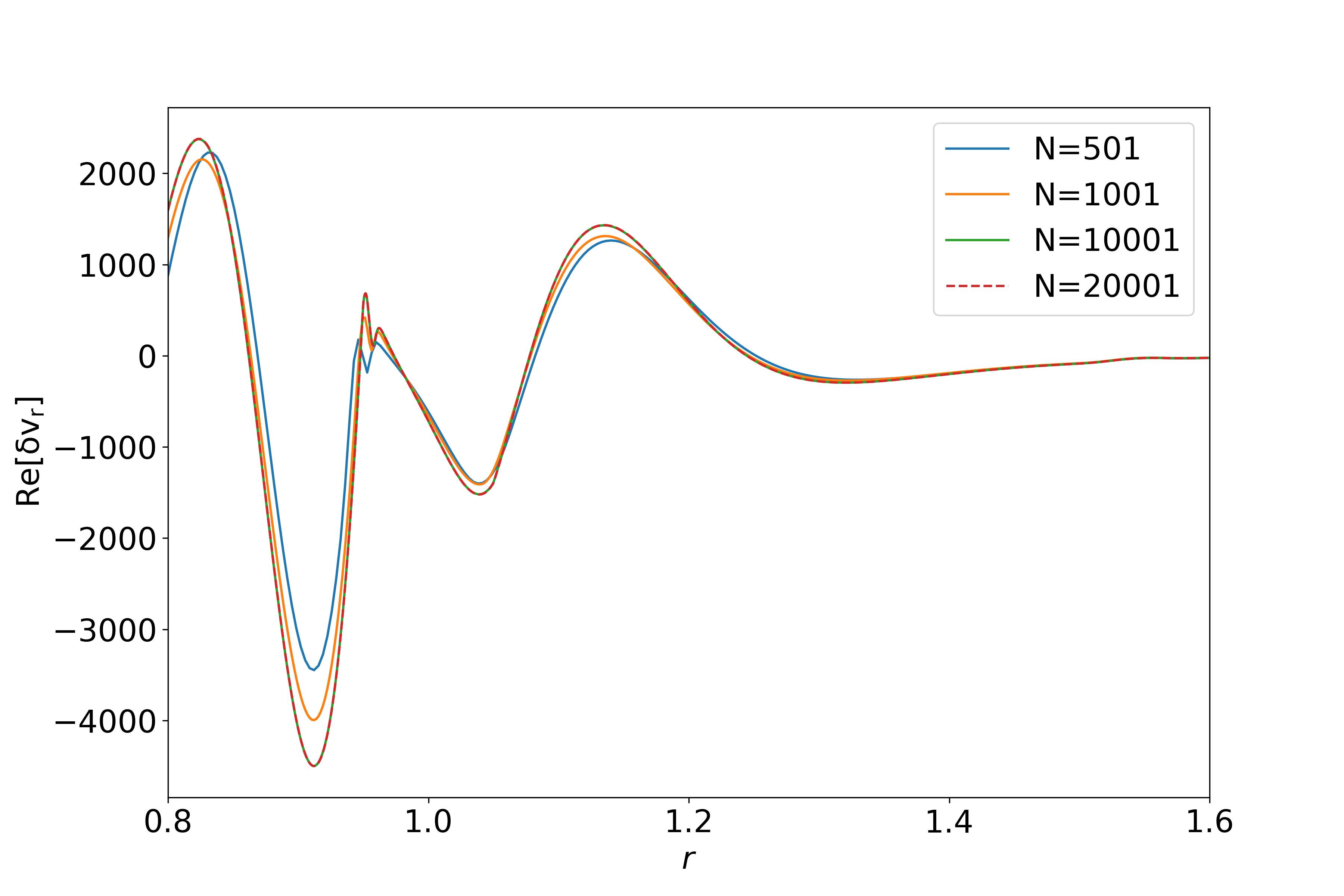}
          }
        \label{}
    \end{minipage}
    \caption{Radial profiles of the real part of $\delta v_r$ for various total grid points N. } 
    \label{Res-pr}
\end{figure*}

The moving average method involves averaging each data point with its neighboring points, which can lead to sharper profiles as resolution increases. 
However, this also raise the issue of the correlation of the global modes with the resolution.
In this investigation, we set the number of grid points used for averaging to 100 while varying the total number of grid points $N$.
Figure \ref{Res-pr} illustrates the radial profile of $\delta v_r$ for $p$-modes and $r$-modes at $N =501, 1001, 10001$ and $20001$. 
The overall profiles are similar. At the lower resolutions of $N=501$ and $N=1001$, the results are significantly different. However, at the higher resolutions of $N=10001$ and $N=20001$, there is no difference in the results, suggesting that the results converge as the resolution increases.

\section{Derivation of dispersion relations}
\label{AppendixA}
In this section, we derive the dispersion relation for MHD density waves in the disk.  
To simplify the analysis, we consider a rigidly rotating disk with a constant angular velocity $\Omega$, resulting in the epicyclic frequency given by $\kappa^2 = 4\Omega^2$.
Next, we assume spatial uniformity across the disk for physical quantities including density $\rho$, pressure $P$, and azimuthal magnetic field $B_{\phi}$. 
Employing the WKB approximation, we can replace $\partial/\partial r$ with $ik_r$. 
Additionally, using the local approximation, we can neglect the curvature term with $r^{-1}$ and replace the subscripts $(r,\phi,z)$ with $(x,y,z)$.
By setting $k_z = 0$ and starting from equations \eqref{dmass} - \eqref{denergy}, we derive the following matrix equation
\begin{equation}\label{Matix_MHD}
    \begin{bmatrix}
        -i\sigma\frac{1}{\rho} & 0 & ik_x & ik_y & 0 & 0 & 0 & 0 \\
        0 & -ik_x & i\sigma & \kappa & 0 & \frac{1}{\rho}ik_yB_y & 0 & 0 \\
        0 & ik_y & \kappa & -i\sigma & 0 & 0 & -\frac{1}{\rho}ik_y B_y & 0 \\
        0 & 0 & 0 & 0 & \sigma & 0 & 0 & \frac{1}{\rho}k_y B_y \\
        0 & 0 & k_y B_y & 0 & 0 & \sigma & 0 & 0 \\
        0 & 0 & ik_xB_y & 0 & 0 & 0 & -i\sigma & 0 \\
        0 & 0 & 0 & 0 & k_yB_y & 0 & 0 & \sigma \\
        \frac{c_s^2}{\rho} & -1 &  0 & 0 & 0 & 0 &\frac{1}{\rho}B_y & 0 \\
    \end{bmatrix}
    \begin{bmatrix}
        \delta \rho  \\
        \Psi  \\
        \delta v_x  \\
        \delta v_y   \\
        \delta v_z   \\
        \delta B_x   \\
        \delta B_y   \\
        \delta B_z   \\
    \end{bmatrix}
     = 0.
\end{equation}

According to the principles of linear algebra, a homogeneous system of linear equations has nontrivial solutions if the determinant of its coefficient matrix is equal to zero. 
Consequently, we acquire the dispersion relation 
\begin{align}
\sigma(\sigma^2 - k_y^2 c_a^2) \nonumber \left\{
\sigma^4
-\left[(c_s^2 + c_a^2)(k_x^2 + k_y^2)+\kappa^2 \right]\sigma^2 \right. \nonumber \left. + (k_x^2 + k_y^2)k_y^2 c_a^2 c_s^2
\right\} = 0.
\end{align}
Notably, $\sigma = 0$ corresponds to the marginal entropy wave \citep{2004prma.book.....G}, which bears no physical significance here, and thus can be disregarded.
An equivalent outcome can also be obtained by initiating the analysis from equations \eqref{leq1}-\eqref{coe22}.

\section{Interpretation of Global Modes}
\label{Appendix_Dispersion}
\begin{figure*}[t]
    \centering
    \begin{minipage}[t]{0.45\textwidth}
        \centering
            \subfigure[]
        {
        \centering
        \includegraphics[width=\linewidth]{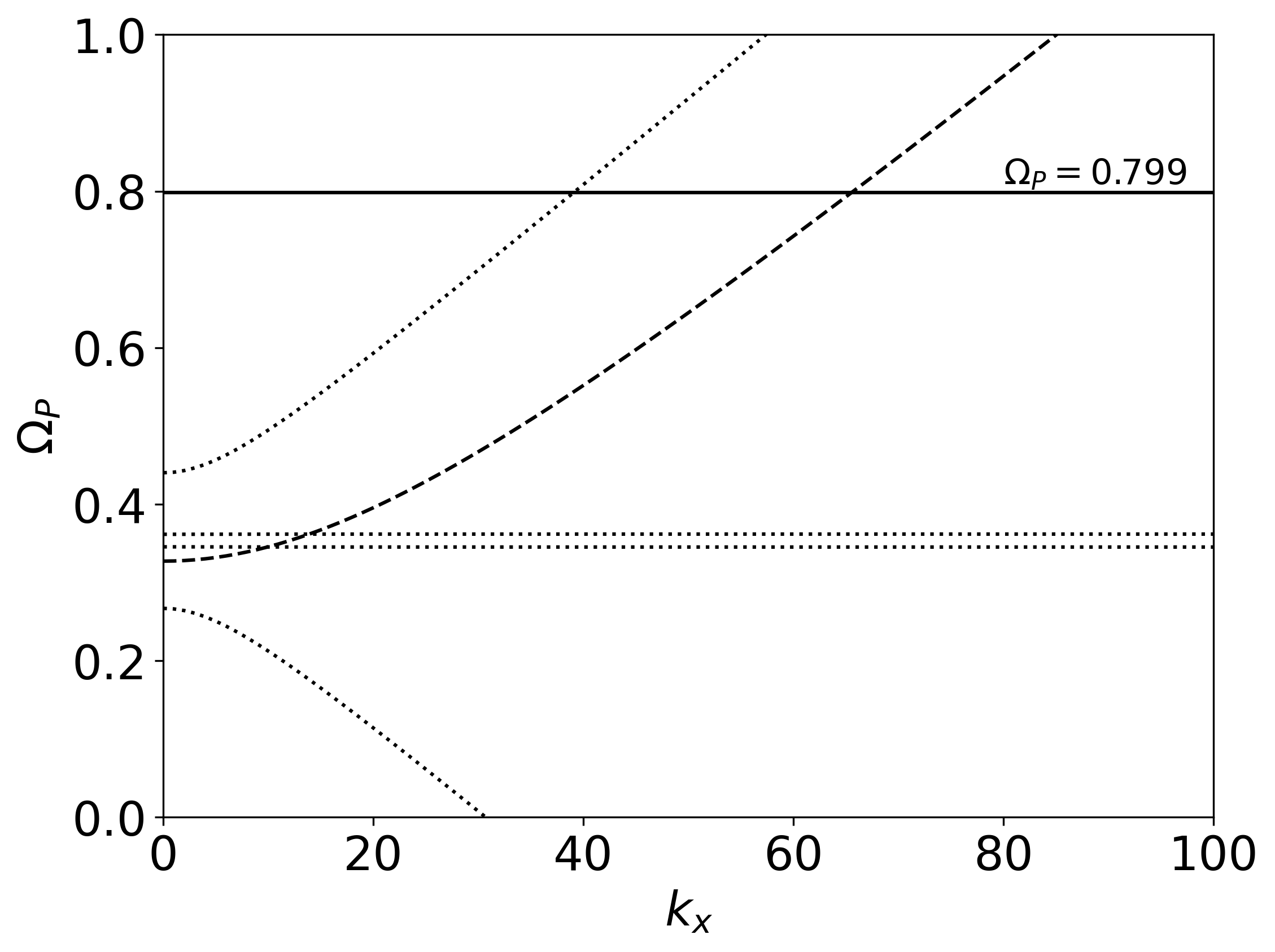}
          }
        \label{Explantion:a}
    \end{minipage}
    \hfill
    \begin{minipage}[t]{0.45\textwidth}
        \centering
            \subfigure[]
        {
        \centering
        \includegraphics[width=\linewidth]{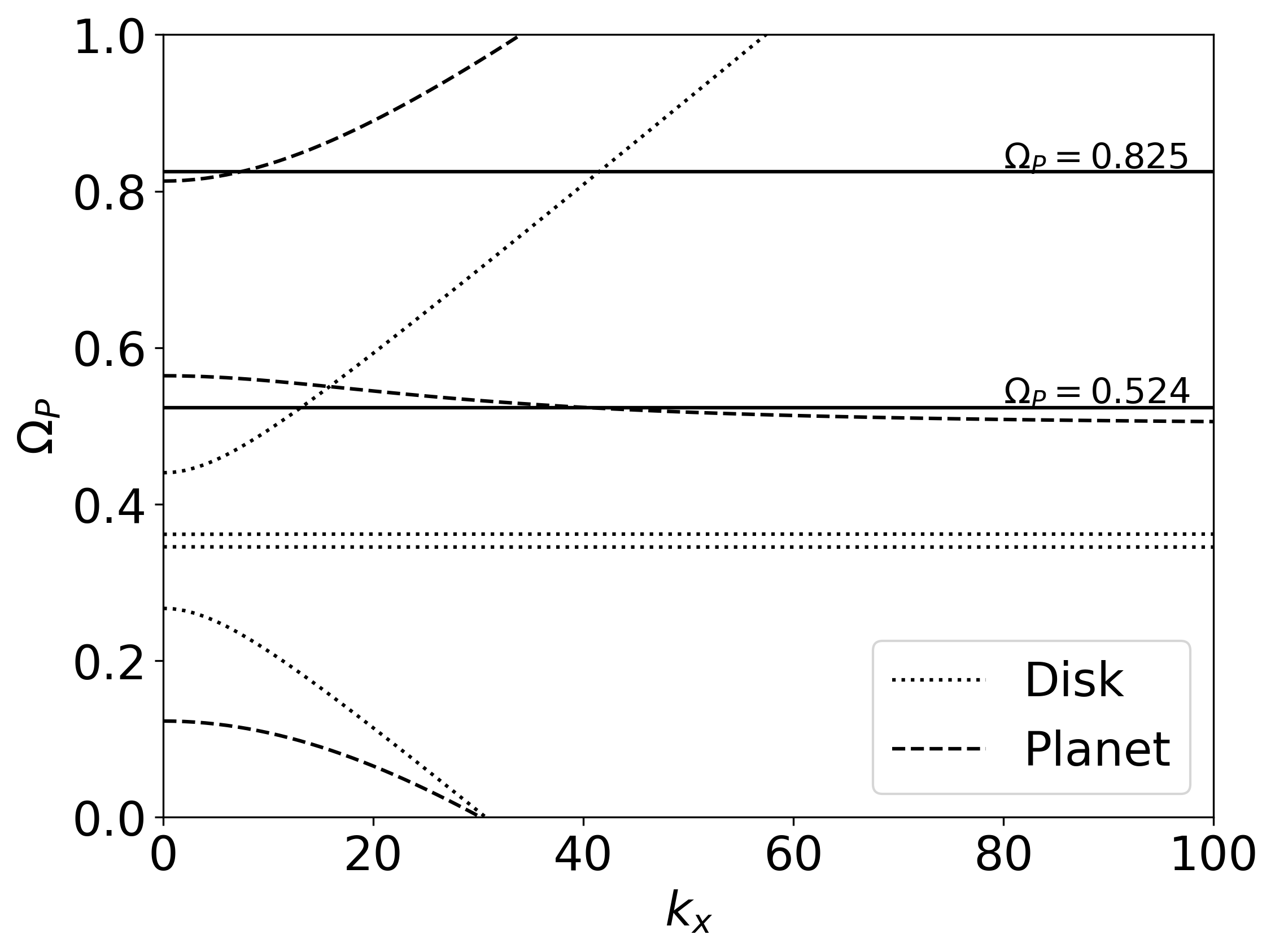}
          }
        \label{Explantion:b}
    \end{minipage}
    \caption{Schematic diagram of $\Omega_{\rm P} - k_x$, corresponding to parameters: (a) $m= 15, \delta r = 0.05, \Omega_0 = 0, \mathcal{M} =6, \lambda = 0.2, k_z =0 $. (b) $m= 15, \delta r = 0.05, \Omega_0 = 0.5, \mathcal{M} =6, \lambda = 0.2, k_z =0 $.  The dashed lines are the dispersion relation for the planetary atmosphere and the dotted lines are the dispersion relation for the disk. $\Omega_{\rm P}$ calculated by our program are represented by black lines.} 
    \label{Explantion}
\end{figure*}

In this section we examine the dispersion relation in detail and use it as a basis for discussing how the global modes are selected.
As we explained in Paper I, since the global mode satisfies both the inner and outer dispersion relations, we simply plot them on the same graph. Figure \ref{Explantion} show the dispersion relations of two cases under parameters: (a) $m= 15, \delta r = 0.05, \Omega_0 = 0, \mathcal{M} =6, \lambda = 0.2, k_z =0 $. (b) $m= 15, \delta r = 0.05, \Omega_0 = 0.5, \mathcal{M} =6, \lambda = 0.2, k_z =0 $. 
For a fixed value of $\Omega_{\rm P}$ , if it can intersect both the inner dispersion relation curve and the outer dispersion relation curve, then it can form a global mode. Otherwise, it cannot form a global mode and this frequency is forbidden. 
Our calculations give the results of each case, including the growth rates $\rm Im[\omega]$, the wavenumber $k_1$ at $r_1$ and the wavenumber $k_2$ at $r_2$.
The former case is $\Omega_{\rm P} = 0.799$, which corresponds to the results $\rm Im[\omega] = 0.173$, $k_1 = -63.7$ and $k_2 =39.1$. The latter case has two results, one is $\Omega_{\rm P} = 0.825$ which corresponds to $\rm Im[\omega] = 0.130 $, $k_1 = -9.2$ and $k_2 =41.5$. Another one is $\Omega_{\rm P} = 0.524$, corresponding to $\rm Im[\omega] = 0.179 $, $k_1 = -35.7$ and $k_2 =13.0$.
We plot the pattern speed calculated by our program for each case. 
Examining the values of $k_x$ at the intersection points, we can find that there is only a small deviation from the calculated results, so our analysis is valid.

\begin{figure*}[t]
    \centering
    \begin{minipage}[t]{0.45\textwidth}
        \centering
            \subfigure[]
        {
        \centering
        \includegraphics[width=\linewidth]{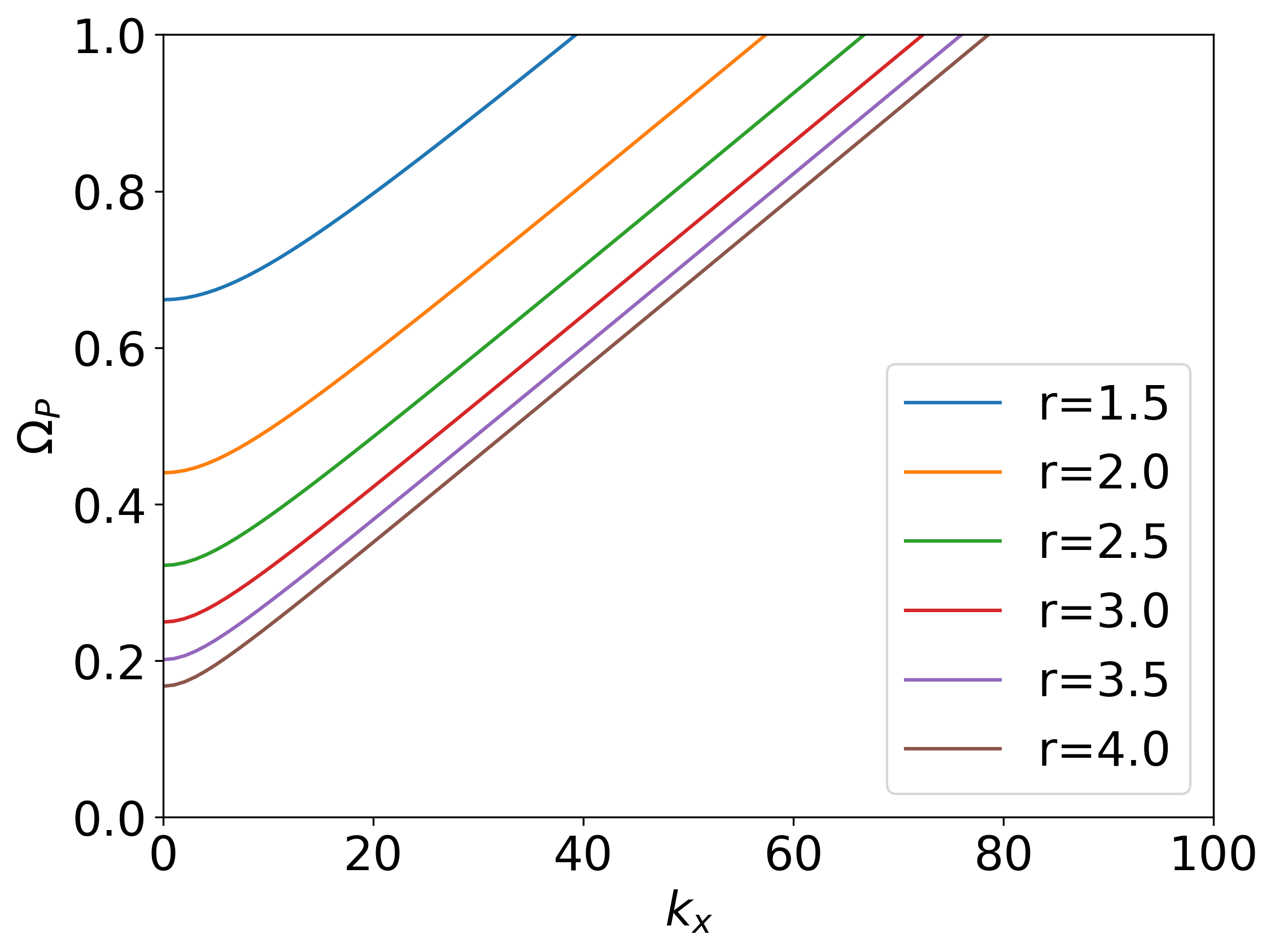}
          }
        \label{Dispersion_radius:a}
    \end{minipage}
    \hfill
    \begin{minipage}[t]{0.45\textwidth}
        \centering
            \subfigure[]
        {
        \centering
        \includegraphics[width=\linewidth]{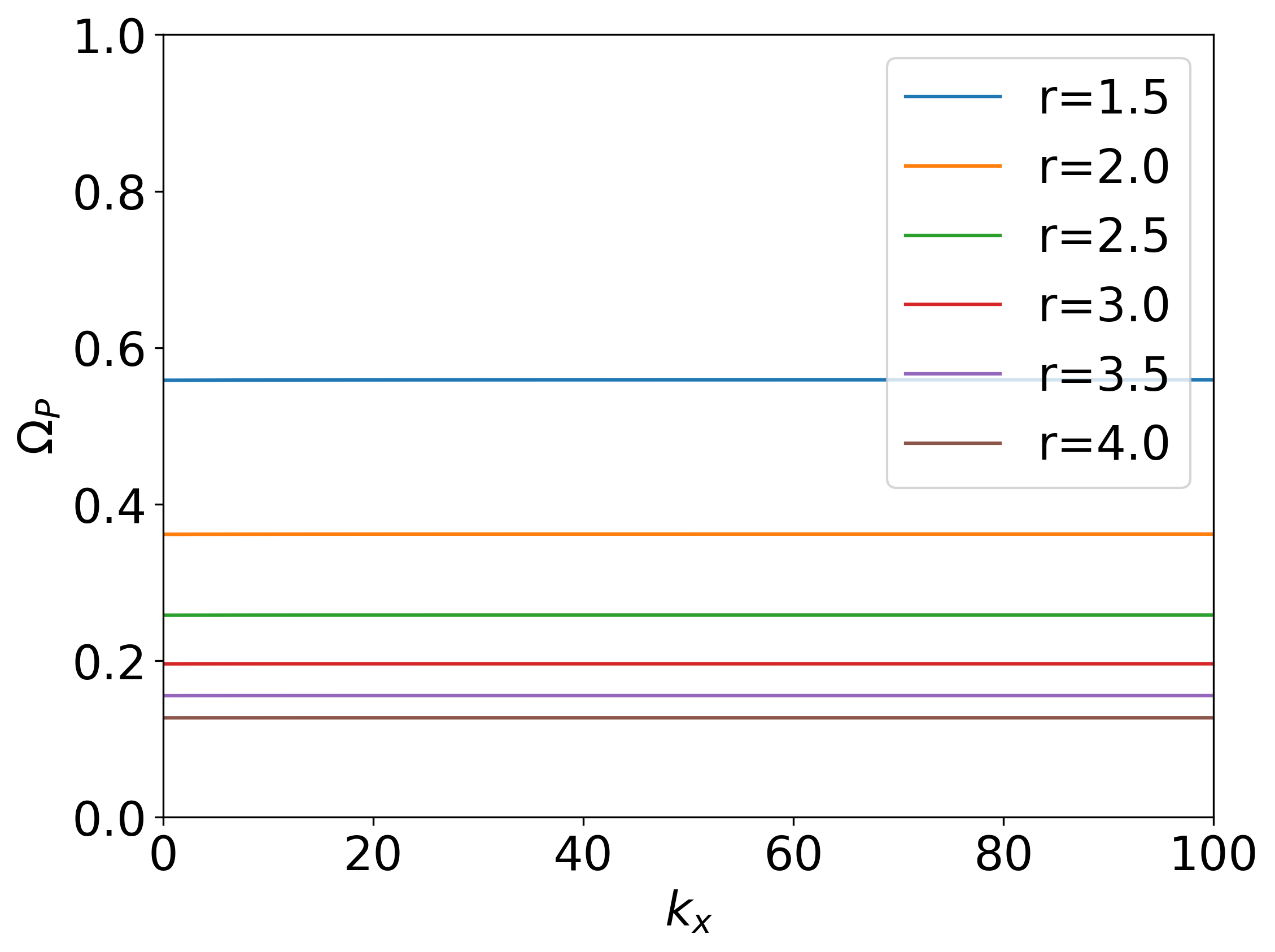}
          }
        \label{Dispersion_radius:b}
    \end{minipage}
    \caption{Dispersion relations for disks of different radii. Panel (a) corresponds to the upper fast MHD density waves. Panel (b) corresponds to the upper slow MHD density waves.} 
    \label{Dispersion_radius}
\end{figure*}

As shown in Figure \ref{Explantion}, a new branch of waves in the disk, namely slow MHD density waves, appears compared to the hydrodynamic case. These waves have a lower frequency if we compare them with the Keplerian rotation speed at $r_2$ ($\Omega_{\rm K} = 0.35$).
It seems like that slow MHD density waves can also form a kind of global modes because they intersect with the dispersion relation of planetary atmosphere. However, we did not find any new global modes formed by slow MHD density waves during our calculations.
This is because this kind of waves are closely related to the local Keplerian speed. If we change the radius, we see that the frequency also changes for different radii, as shown in Figure \ref{Dispersion_radius}(b).
As a result, maintaining a consistent $\Omega_{\rm P}$ across all radii becomes impossible, so slow MHD density waves cannot form a global mode.
However, although fast MHD density waves also vary with radius, we can still find a consistent $\Omega_{\rm P}$ such that it satisfies the dispersion relation for all radii in the disk, as shown in Figure \ref{Dispersion_radius}(a).
In light of this, we expect that the outer boundary conditions should be the fast MHD density waves. This explanation clarifies why \cite{2013ApJ...770...68B} did not identify new magnetosonic modes but rather observed analogous acoustic modes.

\section*{Acknowledgement}
We thank the anonymous referee for helpful suggestions that have greatly improved this paper.
This work has been supported by the National SKA Program of China (Grant No. 2022SKA 0120101) and the National Key R$\&$D Program of China (No. 2020YFC2201200), the science research grants from the China Manned Space Project (No. CMSCSST-2021-B09, CMSCSST-2021-B12 and CMS-CSST-2021-A10), and opening fund of State Key Laboratory of Lunar and Planetary Sciences (Macau University of Science and Technology) (Macau FDCT Grant No. SKL-LPS(MUST)-2021-2023). C.Y. has been supported by the National Natural Science Foundation of China (Grant Nos. 11521303, 11733010, 11873103, and 12373071).

\bibliography{sample631}{}
\bibliographystyle{aasjournal}



\end{CJK*}
\end{document}